%
%
\input harvmac
\input epsf

\def\figin{\epsfcheck\figin}\def\figins{\epsfcheck\figins}
\def\epsfcheck{\ifx\epsfbox\UnDeFiNeD
\message{(NO epsf.tex, FIGURES WILL BE IGNORED)}
\gdef\figin##1{\vskip2in}\gdef\figins##1{\hskip.5in}
\else\message{(FIGURES WILL BE INCLUDED)}%
\gdef\figin##1{##1}\gdef\figins##1{##1}\fi}
\def\DefWarn#1{}
\def\figinsert{\goodbreak\topinsert}
\def\ifig#1#2#3#4{\DefWarn#1\xdef#1{fig.~\the\figno}
\writedef{#1\leftbracket fig.\noexpand~\the\figno}%
\figinsert\figin{\centerline{\epsfxsize=#3mm \epsfbox{#2}}}
\bigskip\medskip\centerline{\vbox{\baselineskip12pt
\advance\hsize by -1truein\noindent\footnotefont{\sl Fig.~\the\figno:}\sl\ #4}}
\bigskip\endinsert\noindent\global\advance\figno by1}

\def\C{{\bf C}}
\def\R{{\bf R}}
\def\Z{{\bf Z}}

\def\a{\alpha}
\def\b{\beta}

\def\e{\epsilon}

\def\l{\lambda}
\def\L{\Lambda}

\def\F{\Phi}
\def\w{\omega}

\def\P{{\bf P}}

\def\d{\partial}
\def\dbar{{\overline\partial}}
\def\inv{^{-1}}
\def\Tr{{\rm Tr}}
\def\hf{{1\over 2}}
\def\cO{{\cal O}}
\def\cC{{\cal C}}
\def\cF{{\cal F}}
\def\cN{{\cal N}}
\def\cW{{\cal W}}
\def\cG{{\cal G}}

\def\({\bigl(}
\def\){\bigr)}
\def\<{\langle\,}
\def\>{\,\rangle}

\lref\thooft{
G.~'t Hooft, ``A Planar Diagram Theory For Strong Interactions,''
Nucl.\ Phys.\ B {\bf 72}, 461 (1974).}

\lref\mm{
P.~Ginsparg and G.~W.~Moore,
``Lectures On 2-D Gravity And 2-D String Theory,''
arXiv:hep-th/9304011.}

\lref\mmm{
P.~Di Francesco, P.~Ginsparg and J.~Zinn-Justin,
``2-D Gravity and random matrices,''
Phys.\ Rept.\  {\bf 254}, 1 (1995)
[arXiv:hep-th/9306153]}

\lref\kontsevich{M.~Kontsevich,
``Intersection Theory On The Moduli Space Of Curves And The Matrix
Airy Function,'' Commun.\ Math.\ Phys.\ {\bf 147}, 1 (1992).}

\lref\wittentop{E.~Witten,
``On The Structure Of The Topological Phase Of Two-Dimensional
Gravity,'' Nucl.\ Phys.\ B {\bf 340}, 281 (1990)}

\lref\bfss{T.~Banks, W.~Fischler, S.~H.~Shenker and L.~Susskind,
``M theory as a matrix model: A conjecture,'' Phys.\ Rev.\ D {\bf 55},
5112 (1997) [arXiv:hep-th/9610043].}

\lref\adscft{
O.~Aharony, S.~S.~Gubser, J.~M.~Maldacena, H.~Ooguri and Y.~Oz,
``Large $N$ field theories, string theory and gravity,'' Phys.\ Rept.\
{\bf 323}, 183 (2000) [arXiv:hep-th/9905111].  }

\lref\ghv{D. Ghoshal and C. Vafa, ``$c=1$ string as the topological theory
of the conifold,'' Nucl.\ Phys.\ B {\bf 453}, 121 (1995)
[arXiv:hep-th/9506122].}

\lref\klmkv{S. Kachru, A. Klemm, W. Lerche, P. Mayr, C. Vafa,
``Nonperturbative Results on the Point Particle Limit of $N=2$
Heterotic String Compactifications,'' Nucl.\ Phys.\ B {\bf 459}, 537
(1996) [arXiv:hep-th/9508155].}

\lref\klw{A. Klemm, W. Lerche, P. Mayr, C.Vafa, N. Warner,
``Self-Dual Strings and $N=2$ Supersymmetric Field Theory,''
 Nucl.\ Phys. \ B {\bf 477}, 746 (1996)
[arXiv:hep-th/9604034].}

\lref\kmv{S. Katz, P. Mayr, C. Vafa,
``Mirror symmetry and Exact Solution of 4D $N=2$ Gauge Theories I,''
Adv.\ Theor.\ Math.\ Phys. {\bf 1}, 53 (1998)
[arXiv:hep-th/9706110].}

\lref\gv{R.~Gopakumar and C.~Vafa,
``On the gauge theory/geometry correspondence,'' Adv.\ Theor.\ Math.\
Phys.\ {\bf 3}, 1415 (1999) [arXiv:hep-th/9811131].  }

\lref\edel{J.D. Edelstein, K. Oh and R. Tatar, ``Orientifold,
geometric transition and large $N$ duality for $SO/Sp$ gauge
theories,'' JHEP {\bf 0105}, 009 (2001) [arXiv:hep-th/0104037].}

\lref\dasg{K. Dasgupta, K. Oh and R. Tatar, ``Geometric transition, large
$N$ dualities and MQCD dynamics,'' Nucl. Phys.  B {\bf 610}, 331
(2001) [arXiv:hep-th/0105066]\semi -----, ``Open/closed string
dualities and Seiberg duality from geometric transitions in
M-theory,'' [arXiv:hep-th/0106040]\semi -----, ``Geometric transition
versus cascading solution,'' JHEP {\bf 0201}, 031 (2002)
[arXiv:hep-th/0110050].}

\lref\hv{K. Hori and C. Vafa,
``Mirror Symmetry,'' [arXiv:hep-th/0002222].}

\lref\hiv{K. Hori, A. Iqbal and C. Vafa,
``D-Branes And Mirror Symmetry,'' [arXiv:hep-th/0005247].}

\lref\vaug{C.~Vafa,
``Superstrings and topological strings at large $N$,''
J.\ Math.\ Phys.\  {\bf 42}, 2798 (2001)
[arXiv:hep-th/0008142].}

\lref\civ{
F.~Cachazo, K.~A.~Intriligator and C.~Vafa,
``A large $N$ duality via a geometric transition,''
Nucl.\ Phys.\ B {\bf 603}, 3 (2001)
[arXiv:hep-th/0103067].}

\lref\ckv{
F.~Cachazo, S.~Katz and C.~Vafa,
``Geometric transitions and $N = 1$ quiver theories,''
arXiv:hep-th/0108120.}

\lref\cfikv{
F.~Cachazo, B.~Fiol, K.~A.~Intriligator, S.~Katz and C.~Vafa, ``A
geometric unification of dualities,'' Nucl.\ Phys.\ B {\bf 628}, 3
(2002) [arXiv:hep-th/0110028].}

\lref\cv{
F.~Cachazo and C.~Vafa, ``$N=1$ and $N=2$ geometry from fluxes,''
arXiv:hep-th/0206017.}

\lref\ov{
H.~Ooguri and C.~Vafa, ``Worldsheet derivation of a large $N$ duality,''
arXiv:hep-th/0205297.}

\lref\av{
M.~Aganagic and C.~Vafa, ``$G_2$ manifolds, mirror symmetry, and
geometric engineering,'' arXiv:hep-th/0110171.}

\lref\digra{
D.~E.~Diaconescu, B.~Florea and A.~Grassi, ``Geometric transitions and
open string instantons,'' arXiv:hep-th/0205234.}

\lref\amv{
M.~Aganagic, M.~Marino and C.~Vafa, ``All loop topological string
amplitudes from Chern-Simons theory,'' arXiv:hep-th/0206164.}

\lref\dfg{
D.~E.~Diaconescu, B.~Florea and A.~Grassi, ``Geometric transitions,
del Pezzo surfaces and open string instantons,''
arXiv:hep-th/0206163.}

\lref\kkl{
S.~Kachru, S.~Katz, A.~E.~Lawrence and J.~McGreevy, ``Open string
instantons and superpotentials,'' Phys.\ Rev.\ D {\bf 62}, 026001
(2000) [arXiv:hep-th/9912151].}

\lref\bcov{
M.~Bershadsky, S.~Cecotti, H.~Ooguri and C.~Vafa, ``Kodaira-Spencer
theory of gravity and exact results for quantum string amplitudes,''
Commun.\ Math.\ Phys.\ {\bf 165}, 311 (1994) [arXiv:hep-th/9309140].
}

\lref\witcs{
E.~Witten, ``Chern-Simons gauge theory as a string theory,''
arXiv:hep-th/9207094.  }

\lref\naret{I. Antoniadis, E. Gava, K.S. Narain, T.R. Taylor,
``Topological Amplitudes in String Theory,'' Nucl.\ Phys.\ B\ {\bf
413}, 162 (1994) [arXiv:hep-th/9307158].}

\lref\witf{E. Witten,
``Solutions Of Four-Dimensional Field Theories Via M Theory,'' Nucl.\
Phys.\ B\ {\bf 500}, 3 (1997) [arXiv:hep-th/9703166].}

\lref\shenker{
S.~H.~Shenker, ``The Strength Of Nonperturbative Effects In String
Theory,'' in Proceedings Cargese 1990, {\it Random surfaces and
quantum gravity}, 191--200.  }

\lref\berwa{
M.~Bershadsky, W.~Lerche, D.~Nemeschansky and N.~P.~Warner, ``Extended
$N=2$ superconformal structure of gravity and W gravity coupled to
matter,'' Nucl.\ Phys.\ B {\bf 401}, 304 (1993)
[arXiv:hep-th/9211040]}

\lref\loopequation{
G.~Akemann, ``Higher genus correlators for the Hermitian matrix model
with multiple cuts,'' Nucl.\ Phys.\ B {\bf 482}, 403 (1996)
[arXiv:hep-th/9606004].  }

\lref\wiegmann{P.B. Wiegmann and A. Zabrodin, ``Conformal maps
and integrable hierarchies,'' arXiv:hep-th/9909147.}

\lref\cone{R.~Dijkgraaf and C.~Vafa, to appear.}

\lref\kazakov{
S.~Y.~Alexandrov, V.~A.~Kazakov and I.~K.~Kostov, ``Time-dependent
backgrounds of 2D string theory,'' arXiv:hep-th/0205079.}

\lref\dj{
S.~R.~Das and A.~Jevicki, ``String Field Theory And Physical
Interpretation Of $D=1$ Strings,'' Mod.\ Phys.\ Lett.\ A {\bf 5}, 1639
(1990).}

\lref\givental{
A.B.~Givental, ``Gromov-Witten invariants and quantization of
 quadratic hamiltonians,'' arXiv:math.AG/0108100.}

\lref\op{
A.~Okounkov and R.~Pandharipande, ``Gromov-Witten theory, Hurwitz
theory, and completed cycle,'' arXiv:math.AG/0204305.}

\lref\dijk{ R.~Dijkgraaf, ``Intersection theory, integrable hierarchies and
topological field theory,'' in Cargese Summer School on {\it New
Symmetry Principles in Quantum Field Theory} 1991,
[arXiv:hep-th/9201003].}

\lref\gw{
D.~J.~Gross and E.~Witten, ``Possible Third Order Phase Transition In
The Large $N$ Lattice Gauge Theory,'' Phys.\ Rev.\ D {\bf 21}, 446
(1980).}

\lref\sw{
N.~Seiberg and E.~Witten, ``Electric-magnetic duality, monopole
condensation, and confinement in $N=2$ supersymmetric Yang-Mills
theory,'' Nucl.\ Phys.\ B {\bf 426}, 19 (1994) [Erratum-ibid.\ B {\bf
430}, 485 (1994)] [arXiv:hep-th/9407087].}

\lref\kkv{
S.~Katz, A.~Klemm and C.~Vafa, ``Geometric engineering of quantum
field theories,'' Nucl.\ Phys.\ B {\bf 497}, 173 (1997)
[arXiv:hep-th/9609239].}

\lref\taylor{
W.~I.~Taylor, ``D-brane field theory on compact spaces,'' Phys.\
Lett.\ B {\bf 394}, 283 (1997) [arXiv:hep-th/9611042].}

\lref\kmmms{
S.~Kharchev, A.~Marshakov, A.~Mironov, A.~Morozov and S.~Pakuliak,
``Conformal matrix models as an alternative to conventional
multimatrix models,'' Nucl.\ Phys.\ B {\bf 404}, 717 (1993)
[arXiv:hep-th/9208044].}

\lref\dv{
R.~Dijkgraaf and C.~Vafa, ``Matrix models, topological strings, and
supersymmetric gauge theories,'' arXiv:hep-th/0206255.}

\lref\kostov{
I.~K.~Kostov, ``Gauge invariant matrix model for the A-D-E closed
strings,'' Phys.\ Lett.\ B {\bf 297}, 74 (1992)
[arXiv:hep-th/9208053].  }

\lref\ot{
K.~h.~Oh and R.~Tatar, ``Duality and confinement in $N=1$
supersymmetric theories from geometric transitions,''
arXiv:hep-th/0112040.}

\lref\ovknot{
H.~Ooguri and C.~Vafa, ``Knot invariants and topological strings,''
Nucl.\ Phys.\ B {\bf 577}, 419 (2000) [arXiv:hep-th/9912123].  }

\lref\dvv{
R.~Dijkgraaf, H.~Verlinde and E.~Verlinde, ``Loop Equations And
Virasoro Constraints In Nonperturbative 2-D Quantum Gravity,'' Nucl.\
Phys.\ B {\bf 348}, 435 (1991).}

\lref\kawai{
M.~Fukuma, H.~Kawai and R.~Nakayama, ``Continuum Schwinger-Dyson
Equations And Universal Structures In Two-Dimensional Quantum
Gravity,'' Int.\ J.\ Mod.\ Phys.\ A {\bf 6}, 1385 (1991).}

\lref\nekrasov{
N.~A.~Nekrasov, ``Seiberg-Witten prepotential from instanton
counting,'' arXiv:hep-th/0206161.}

\lref\ki{
I.~K.~Kostov, ``Bilinear functional equations in 2D quantum gravity,''
in Razlog 1995, {\it New trends in quantum field theory}, 77--90,
[arXiv:hep-th/9602117].}

\lref\kii{
I.~K.~Kostov, ``Conformal field theory techniques in random matrix
models,'' arXiv:hep-th/9907060.}

\lref\morozov{
A.~Morozov, ``Integrability And Matrix Models,'' Phys.\ Usp.\ {\bf
37}, 1 (1994) [arXiv:hep-th/9303139].}

\lref\fo{
H.~Fuji and Y.~Ookouchi, ``Confining phase superpotentials for SO/Sp
gauge theories via geometric transition,'' arXiv:hep-th/0205301.  }

\lref\marinor{M. Marino, ``Chern-Simons theory, matrix integrals, and
perturbative three-manifold invariants,'' [arXiv:hep-th/0207096].}

\lref\agnt{
I.~Antoniadis, E.~Gava, K.~S.~Narain and T.~R.~Taylor, ``Topological
amplitudes in string theory,'' Nucl.\ Phys.\ B {\bf 413}, 162 (1994)
[arXiv:hep-th/9307158].}

\lref\vw{
C.~Vafa and E.~Witten, ``A Strong coupling test of S duality,'' Nucl.\
Phys.\ B {\bf 431}, 3 (1994) [arXiv:hep-th/9408074].}

\lref\ps{
J.~Polchinski and M.~J.~Strassler, ``The string dual of a confining
four-dimensional gauge theory,'' arXiv:hep-th/0003136.}

\lref\kkn{
V.~A.~Kazakov, I.~K.~Kostov and N.~A.~Nekrasov, ``D-particles, matrix
integrals and KP hierarchy,'' Nucl.\ Phys.\ B {\bf 557}, 413 (1999)
[arXiv:hep-th/9810035].}

\lref\kpw{
A.~Khavaev, K.~Pilch and N.~P.~Warner, ``New vacua of gauged $N = 8$
supergravity in five dimensions,'' Phys.\ Lett.\ B {\bf 487}, 14
(2000) [arXiv:hep-th/9812035].}

\lref\ks{
S.~Kachru and E.~Silverstein, ``4d conformal theories and strings on
orbifolds,'' Phys.\ Rev.\ Lett.\ {\bf 80}, 4855 (1998)
[arXiv:hep-th/9802183].  }

\lref\bj{
M.~Bershadsky and A.~Johansen, ``Large $N$ limit of orbifold field
theories,'' Nucl.\ Phys.\ B {\bf 536}, 141 (1998)
[arXiv:hep-th/9803249].}

\lref\lnv{
A.~E.~Lawrence, N.~Nekrasov and C.~Vafa, ``On conformal field theories
in four dimensions,'' Nucl.\ Phys.\ B {\bf 533}, 199 (1998)
[arXiv:hep-th/9803015].  }

\lref\bkv{
M.~Bershadsky, Z.~Kakushadze and C.~Vafa, ``String expansion as large
$N$ expansion of gauge theories,'' Nucl.\ Phys.\ B {\bf 523}, 59 (1998)
[arXiv:hep-th/9803076].}

\lref\dvii{
R.~Dijkgraaf and C.~Vafa, ``On geometry and matrix models,''
arXiv:hep-th/0207106.}

\lref\vy{
G.~Veneziano and S.~Yankielowicz, ``An Effective Lagrangian For The
Pure $N=1$ Supersymmetric Yang-Mills Theory,'' Phys.\ Lett.\ B {\bf
113}, 231 (1982).  }

\lref\sinhv{
S.~Sinha and C.~Vafa, ``$SO$ and $Sp$ Chern-Simons at large $N$,''
arXiv:hep-th/0012136.  }

\lref\aahv{B.~Acharya, M.~Aganagic, K.~Hori and C.~Vafa,
``Orientifolds, mirror symmetry and superpotentials,''
arXiv:hep-th/0202208.}

\lref\dorey{N. Dorey, ``An Elliptic Superpotential for Softly
Broken $N=4$ Supersymmetric Yang-Mills,'' JHEP {\bf 9907}, 021 (1999)
[arXiv:hep-th/9906011].}

\lref\ls{
R.~G.~Leigh and M.~J.~Strassler, ``Exactly marginal operators and
duality in four-dimensional $N=1$ supersymmetric gauge theory,'' Nucl.\
Phys.\ B {\bf 447}, 95 (1995) [arXiv:hep-th/9503121].
}

\lref\klyt{
A.~Klemm, W.~Lerche, S.~Yankielowicz and S.~Theisen,
``Simple singularities and $N=2$ supersymmetric Yang-Mills theory,''
Phys.\ Lett.\ B {\bf 344}, 169 (1995)
[arXiv:hep-th/9411048]}

\lref\af{
P.~C.~Argyres and A.~E.~Faraggi, ``The vacuum structure and spectrum
of $N=2$ supersymmetric $SU(n)$ gauge theory,'' Phys.\ Rev.\ Lett.\
{\bf 74}, 3931 (1995) [arXiv:hep-th/9411057].}

\lref\mo{C.~Montonen and D.~I.~Olive,
``Magnetic Monopoles As Gauge Particles?,'' Phys.\ Lett.\ B {\bf 72},
117 (1977).  }

\lref\gvw{S.~Gukov, C.~Vafa and E.~Witten,
``CFT's from Calabi-Yau four-folds,'' Nucl.\ Phys.\ B {\bf 584}, 69
(2000) [Erratum-ibid.\ B {\bf 608}, 477 (2001)]
[arXiv:hep-th/9906070].  }

\lref\tv{
T.~R.~Taylor and C.~Vafa, ``RR flux on Calabi-Yau and partial
supersymmetry breaking,'' Phys.\ Lett.\ B {\bf 474}, 130 (2000)
[arXiv:hep-th/9912152].  }

\lref\mayr{
P.~Mayr, ``On supersymmetry breaking in string theory and its
realization in brane worlds,'' Nucl.\ Phys.\ B {\bf 593}, 99 (2001)
[arXiv:hep-th/0003198].  }

\lref\qfvy{A.~de la Macorra and G.G.~Ross, Nucl.\ Phys.\ B {\bf 404},
321 (1993)\semi C. P. Burgess, J.-P. Derendinger, F. Quevedo,
M. Quiros, ``On Gaugino Condensation with Field-Dependent Gauge
Couplings,'' Annals Phys. {\bf 250}, 193 (1996)
[arXiv:hep-th/9505171].}

\lref\div{
A. D'Adda, A.C. Davis, P. Di Vecchia and P. Salomonson, Nucl. Phys. B
{\bf 222}, 45 (1983).}

\lref\berva{
N. Berkovits and C. Vafa, ``$N=4$ Topological Strings,'' Nucl. Phys. B
{\bf 433} 123 (1995) [arXiv:hep-th/9407190].}

\lref\novikov{
V.~A.~Novikov, M.~A.~Shifman, A.~I.~Vainshtein and V.~I.~Zakharov,
``Instanton Effects In Supersymmetric Theories,'' Nucl.\ Phys.\ B {\bf
229}, 407 (1983).  }

\lref\dk{
N.~Dorey and S.~P.~Kumar, ``Softly-broken $N = 4$ supersymmetry in the
large-$N$ limit,'' JHEP {\bf 0002}, 006 (2000) [arXiv:hep-th/0001103].
}

\lref\ds{
N.~Dorey and A.~Sinkovics, ``$N = 1^*$ vacua, fuzzy spheres and
integrable systems,'' JHEP {\bf 0207}, 032 (2002)
[arXiv:hep-th/0205151].  }

\lref\ads{
I.~Affleck, M.~Dine and N.~Seiberg, ``Supersymmetry Breaking By
Instantons,'' Phys.\ Rev.\ Lett.\ {\bf 51}, 1026 (1983).  }

\lref\bele{
D. Berenstein, V. Jejjala and R. G. Leigh, ``Marginal and Relevant
Deformations of $N=4$ Field Theories and Non-Commutative Moduli Spaces
of Vacua,'' Nucl.Phys. B {\bf 589} 196 (2000) [arXiv:hep-th/0005087].}

\lref\gins{
P. Ginsparg, ``Matrix models of 2d gravity,'' Trieste Lectures (July,
1991), Gava et al., 1991 summer school in H.E.P. and Cosmo.
[arXiv:hep-th/9112013].}

\lref\kostovsix{
I. Kostov, ``Exact Solution of the Six-Vertex Model on a Random
 Lattice,'' Nucl.Phys. B {\bf 575} 513 (2000) [arXiv:hep-th/9911023].}

\lref\hoppe{
J.~Goldstone, unpublished; J.~Hoppe, ``Quantum theory of a massless
relativistic surface,'' MIT PhD thesis, 1982.}

\Title
 {\vbox{ \hbox{hep-th/0208048} \hbox{HUTP-02/A034} \hbox{ITFA-2002-34}
\vskip-10mm
}}
{\vbox{
\centerline{A Perturbative Window into Non-Perturbative Physics}
}}
\centerline{Robbert Dijkgraaf}
\vskip.05in
\centerline{\sl Institute for Theoretical Physics \&}
\centerline{\sl Korteweg-de Vries Institute for Mathematics}
\centerline{\sl University of Amsterdam}
\centerline{\sl 1018 TV Amsterdam, The Netherlands }
\smallskip
\centerline{and}
\smallskip
\centerline{Cumrun Vafa}
\vskip.05in
\centerline{\sl Jefferson Physical Laboratory}
\centerline{\sl Harvard University}
\centerline{\sl Cambridge, MA 02138, USA}
\smallskip

\vskip .1in\centerline{\bf Abstract}

\smallskip

We argue that for a large class of $\cN=1$ supersymmetric gauge
theories the effective superpotential as a function of the glueball
chiral superfield is exactly given by a summation of planar diagrams
of the same gauge theory.  This perturbative computation reduces to a
matrix model whose action is the tree-level superpotential.
For all models that can be embedded in string theory we give a proof
of this result, and we sketch an argument how to derive this more
generally directly in field theory.  These results are obtained
without assuming any conjectured dualities and can be used as a
systematic method to compute instanton effects: the perturbative
corrections up to $n$-th loop can be used to compute up to
$n$-instanton corrections.  These techniques allow us to see many
non-perturbative effects, such as the Seiberg-Witten solutions of
$\cN=2$ theories, the consequences of Montonen-Olive $S$-duality in
$\cN=1^*$ and Seiberg-like dualities for $\cN=1$ theories from a
completely perturbative planar point of view in the same gauge
theory, without invoking a dual description.

\Date{August, 2002}


\newsec{Introduction}

One of the important insights in recent years in theoretical physics
has been the discovery of duality symmetries in gauge theory and string
theory.  In particular we have learned that the dynamics of
supersymmetric gauge theories, in particular non-perturbative effects
at strong coupling, are often captured by some weakly coupled dual
theory. The Montonen-Olive duality of $\cN=4$ \mo\ and the strong
coupling dynamics of $\cN=2$ gauge theory captured by a dual abelian
gauge theory via the Seiberg-Witten geometry \sw\ are among the prime
examples.  These gauge theoretic dualities have been embedded in
string theory dualities where the gauge theory is engineered by
considering a suitbale string background.  In some cases, for example
in the context of Matrix Theory \bfss\ and the AdS/CFT correspondence
\adscft, the string/gauge theory dualities are actually equivalent.

In this paper we wish to argue that in fact essentially all these
gauge theoretic dualities can be seen perturbatively in the {\it same}
gauge theory in the context of computations of exact superpotentials
in $\cN=1$ gauge theories, that are obtained by adding deformations to
the original gauge theory. We will give arguments that these
superpotentials can be computed exactly by summing over all {\it
planar} diagrams of the {\it zero-momentum} modes. Key to this all is
the computation of the effective superpotential as a function of the
glueball chiral superfield $S={1\over 32 \pi^2}\Tr\,W_\alpha
W^{\alpha}$, generalizing the Veneziano-Yankielowicz effective
superpotential for the case of pure Yang-Mills \vy.

That certain quantities can be computed exactly in perturbation theory
has been encountered before in quantum field theories.  In particular
the axial anomaly can be computed exactly by a one loop computation.
However, it is more rare to encounter computable, anomaly-like
quantities at higher loops in perturbation theory.  Such computable
anomaly-like amplitudes were encountered in type II string
perturbation theory \refs{\bcov,\agnt}, where it was shown that
topological string amplitudes on Calabi-Yau threefolds compute F-type
terms in the associated four-dimensional $\cN =2$ supersymmetric
theory.  Furthermore in \bcov\ it was shown that in the context of
type I string theory the topological string computes superpotential
terms involving the glueball superfield at higher loops in open string
perturbation theory, and it was speculated there that this must be
important for aspects of gaugino condensation in $\cN =1$
supersymmetric gauge theories in four dimensions.

These open string computations of the effective superpotential can be
exactly summed if a closed string large $N$ dual can be found.  The
study of large $N$ dualities in the context of topological strings
were initiated in \gv, and embedded in superstrings in \vaug.  This
idea was developed further in a series of papers beginning with \civ\
were highly non-trivial aspects of $\cN=1$ supersymmetric gauge
theories were deduced using this duality \refs{\edel,\dasg
,\ckv,\cfikv,\ot,\fo,\cv}.  More recently it was discovered in
\refs{\dv,\dvii} that the relevant computations in \civ\ are {\it
perturbative} even from the viewpoint of the corresponding gauge
theory and reduce to matrix integrals. The aim of this paper is to
explain the universality of the ideas in \refs{\dv,\dvii} and show its
power in gaining insight into non-perturbative gauge theoretic
phenomena from a perturbative perspective. In particular we explain
how these ideas may be understood from first principles in purely
gauge theoretic terms without appealing to string theory or any other
dualities.

\subsec{A chain of string dualities}

Although we want to stress in the paper that we are making a purely
field-theoretic statement about supersymmetric gauge theories, perhaps
it can be helpful to summarize the chain of arguments in string theory
that have led us to this result.

To actually compute F-terms explicitly in string theory one can make
use of a series of powerful dualities. As a starting point one
typically engineers the $\cN=1$ gauge theory in string theory by a
D-brane configuration, for example by realizing it as a collection of
D5-branes wrapped around two-cycles in a Calabi-Yau geometry in type
IIB theory. This Calabi-Yau is typically a small resolution of a
singular geometry. This open string theory can then have a large $N$
dual closed string realization, where the Calabi-Yau geometry goes
through a so-called geometric transition in which the topology
changes. The two-cycles around which the D-branes were wrapped are
blown down and three-cycles appear. In the process of this transition
the D-branes disappear and emerge as fluxes of a 3-form field strength
$H=H_{RR}+\tau H_{NS}$.

\ifig\cygeometry{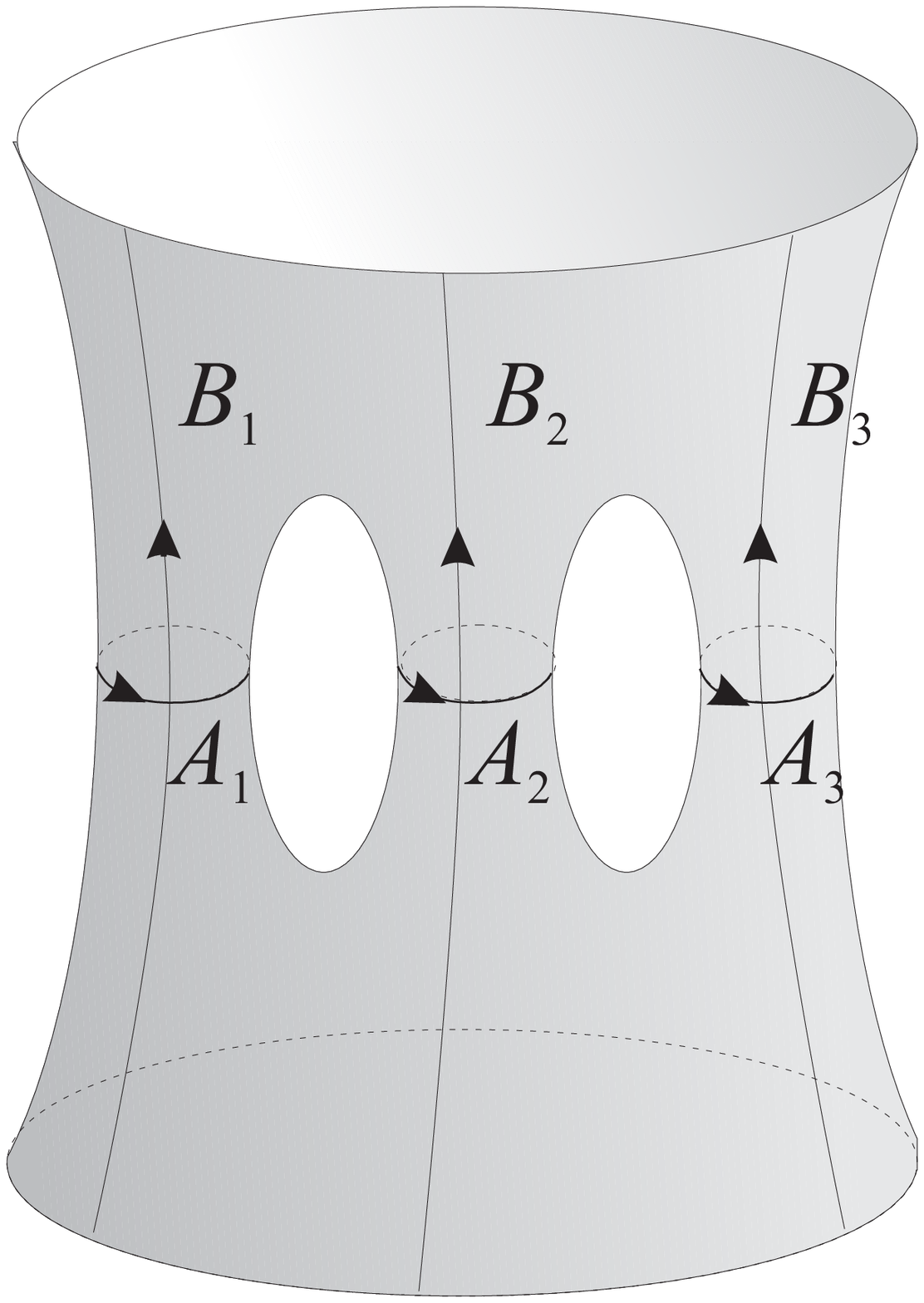}{35}{ 
The Calabi-Yau geometry dual to the gauge theory, with its canonical
homology basis.}  

In the dual closed string theory on the deformed Calabi-Yau there is
an elegant exact expression for the effective superpotential in terms
of the geometry \refs{\gvw,\tv,\mayr}
$$
W_{\rm eff} = \int_X H \wedge \Omega
$$
with $\Omega$ the holomorphic $(3,0)$ form on the Calabi-Yau space
$X$.  Introducing a canonical basis of homology three-cycles
$A_i,B_i$, where the $A$-cycles are typically compact and the
$B$-cycles non-compact (see \cygeometry) this can be written as
$$
W_{\rm eff} = \sum_i \biggl[ \oint_{A_i} H \int_{B_i}\Omega -
\int_{B_i} H \oint_{A_i}\Omega \biggr].
$$  
The periods of the holomorphic three-form give the so-called special
geometry relations
$$
\oint_{A_i} \Omega = 2\pi i S_i,\qquad 
\int_{B_i} \Omega ={\d \cF_0 \over \d S_i}.
$$
The variables $S_i$ can be used as moduli that parametrize the complex
structure of the CY. From the dual gauge theory perspective these
moduli $S_i$ are the glueball condensates that appear when the
strongly coupled gauge theory confines \vaug. To evaluate the
superpotential one further needs the fluxes of the $H$-field through
the cycles
$$
N_i = \oint_{A_i} H,\qquad \tau_i = \int_{B_i} H.
$$
The integers $N_i$ correspond to the number of D-branes (the rank of
the gauge group), and the cut-off dependent variables $\tau_i$
correspond to the bare gauge couplings, $\tau \sim 4\pi i/g^2$.  With
all this notation the final expression for superpotential can be
simply written as
\eqn\flux{
W_{\rm eff}(S) = \sum_i\biggl(N_i {\d\cF_0\over \d S_i}-2\pi i \tau_i
S_i\biggr).
}
The only non-trivial ingredient is the prepotential $\cF_0(S)$ that
encodes the special geometry and that is completely determined by the
CY geometry.

The prepotential $\cF_0(S)$ is well-known to reduce to a computation
in the closed topological string theory, that is purely defined in
terms of the internal CY space.  In fact, the full computation of the
effective superpotential can be done within the topological
string. The same geometric transition that related the closed type IIB
string background to the D-brane system, also relates the closed
topological string to an open topological string propagating in the
background of a collection of branes. These so-called B-branes are
entirely located within the internal CY space---they are simply the
internal part of the original D-branes. The open and closed
topological strings are related by the same large $N$ duality as the
physical theory. The planar diagrams of the open string reproduce the
sphere diagram of the closed string.

But there is an important difference with the physical theory.  In the
topological setting the closed string moduli $S_i$ are expressed in
terms of the number of branes $N_i$ and the string coupling $g_s$
through the 't Hooft couplings \gv
$$
S_i = g_s N_i,
$$ 
defined in the limit $N_i \to \infty$, $g_s\to 0$. So in the
topological context the variables $S_i$ and $N_i$ are {\it not}
independent and the moduli $S_i$ are frozen to the above values. This
should be contrasted with the physical string theory where the
glueball fields $S_i$ are not frozen and independent of $N_i$---the
expectation values of the $S_i$ are set by extremizing the effective
superpotential \flux\ and thus depend on the gauge couplings.

\ifig\chain{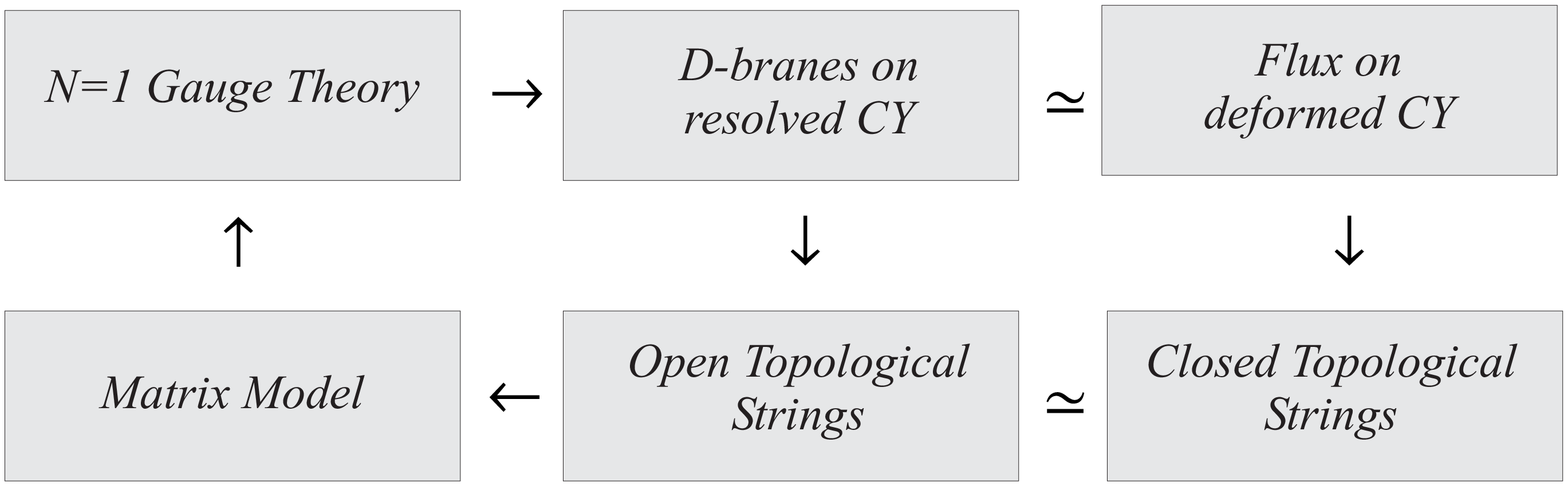}{110}{
The chain of string theory dualities that have been used to compute
effective superpotentials in $\cN=1$ gauge theories.}

Irrespective of the existence of a closed string theory dual defined
by a geometric transition, one can study the open topological string
on itself. Its amplitudes are directly related to F-term computations
in the corresponding D-brane system. In particular its planar diagrams
are directly related to the computation of the superpotential in the
physical gauge theory. The world-volume theory on the B-brane is a
dimensionally reduced version of the holomorphic Chern-Simons gauge
theory introduced in \witcs. As explained in \dv\ under suitable
circumstances this model reduces to only zero-modes, and in this way
the computation reduces to a zero-dimensional field theory---a
generalized matrix model. However, once we have arrived at this point
we have essentially gone through a $360^\circ$ rotation, since we can
relate the diagrams of the matrix model directly to the original gauge
theory, essentially bypassing all the previous dualities---and this
will be the point of view in this paper.  We summarize this chain of
dualities in \chain

\subsec{Outline}

The plan of this paper is as follows: In section 2 we state the
precise relation between ${\cal N}=1$ four-dimensional supersymmetric
gauge theories and matrix models.  In particular this relation shows
that one can use perturbative gauge theory techniques to gain exact
information about highly non-trivial non-perturbative aspects of gauge
theory.  In section 3 we outline the arguments leading to this
statement, both within string theory and in field theory.  In section
4 we discuss the implications of this result for deriving various
exact consequences for gauge theories.  This includes a perturbative
derivation of ${\cal N}=2$ Seiberg-Witten geometry as well as
derivation of ${\cal N}=1$ Seiberg-like dualities.  Moreover we find
that non-perturbative consequences of Olive-Montonen dualities, as
seen in the ${\cal N}=1^*$ deformation of ${\cal N}=4$ Yang-Mills
theory, can also be derived in this setup.  In section 5 we conclude
with some ideas which we are presently pursuing and summarize our main
conclusions.

\newsec{Matrix models and effective superpotentials}

We will formulate a very general conjecture relating exact
superpotentials and matrix models. But before we do this let us first
discuss an example that shows all the features.

\subsec{The canonical example}

The key example to keep in mind is $\cN=1$ $U(N)$ gauge theory with
one adjoint chiral matter field $\F$. This theory is obtained by
softly breaking $\cN=2$ super-Yang-Mills down to $\cN=1$ by means of
the tree-level superpotential
\eqn\tree{
\int d^2\!\theta \; \Tr\, W(\F).
}
Here we take $W(x)$ to be a polynomial of degree $n+1$ in a single
complex variable $x$. Since we do not want to limit ourselves
exclusively to a superpotential of at most degree three, we typically
think of this model as an effective theory obtained by integrating out
other fields in an underlying renormalizable quantum field theory.

Since $W$ has (generically) $n$ isolated critical points at
$x=a_1,\ldots,a_n$, the classical vacua of this theory are determined
by distributing the eigenvalues of the matrix $\F$ over these critical
points. If we choose a partition
$$
N = N_1 + \ldots + N_n,
$$ 
and put $N_i$ eigenvalues at the critical point $a_i$ then we have a
symmetry breaking pattern
$$
U(N) \to U(N_1)\times \cdots \times U(N_n).
$$

The corresponding quantum vacua are described by the appearance of
a gaugino condensate and confinement in the $SU(N_i)$ subgroups of
these $U(N_i)$ factors. The gauge coupling becomes strong at a
dynamically generated scale $\L$ where the gauge group $SU(N_i)$ is
completely broken down and a mass gap is generated. Let
$$
S_i={1\over 32\pi^2}\Tr_{SU(N_i)} W_\a^2
$$
denote the corresponding chiral superfield, whose lowest component is
the gaugino bilinear $\Tr\,\l_\a^2$ that gets a dynamical expectation
value, and whose top component gives the (chiral half) of the
super-Yang-Mills action. The condensate $\< S_i\>$ breaks the
$\Z_{2N_i}$ global symmetry down to $\Z_2$, and thus generates $N_i$
inequivalent vacua. The relevant physical quantities to compute in
these confining vacua are the values of the gaugino condensate and the
tensions of the domain walls interpolating between the different
vacua, that can be expressed as the differences of the values of the
effective superpotential in these confining vacua. Of course, all
these quantities will be highly non-trivial functions of the coupling
constants of the bare superpotential $W_{\rm tree}(\F)$, and it is
these functions that we are after.

The effect of gaugino condensation is elegantly described by an
effective superpotential which is a function of the chiral superfields
$S_i$
$$
\int d^2\!\theta\,W_{\rm eff}(S_i).
$$
Minimizing this action with respect to the variables $S_i$ then
describes the vacuum structure of the theory and determines the
computable physical quantities. For a pure $SU(N)$ gauge theory this
effective superpotential takes the Veneziano-Yankielowicz form
\vy\ (up to non-universal terms that can be absorbed in the cut-off 
scale $\L_0$)
\eqn\vypot{
W_{\rm eff}(S)=NS\log (S/\L_0^3)-2\pi i \tau S, }
with $\tau$ the usual combination of the bare gauge coupling $g$ and
the theta angle
$$
\tau ={\theta\over 2\pi} + {4\pi i \over g^2}.
$$
Minimizing the action with respect to $S$ gives the value of the
condensate in the $N$ distinct vacua labeled by $k=0,1,\ldots,N-1$
\eqn\gaugino{
\<S\> \sim \L^3 e^{2\pi i k/N},
}
with $\L$ the dynamically generated scale.  The gaugino condensate is
interpretated as a non-perturbative effect due to a fractional
instanton of charge $1/N$. Equivalently, the $N$-point function of $S$
is constant and saturated by a one-instanton field configuration, $\<
S^N\>\sim e^{2\pi i \tau}$ \novikov.

The $S\log S$ term in the action \vypot\ strongly suggests a
derivation of the superpotential by a perturbative---in fact
one-loop---Coleman-Weinberg-like computation. As such it should
perhaps be compared to the analogous one loop computation of the term
$N\Sigma\log\Sigma $ in the two-dimensional linear sigma model for the
supersymmetric ${\bf CP}^{N-1}$ sigma model \div.  The corresponding
argument in the four-dimensional case uses the anomalous R-symmetry
\vy: the anomalous axial $U(1)$ symmetry implies that rotating
$$
S\rightarrow e^{2\pi i}S
$$ 
shifts the theta angle by $2\pi N$ which implies that the
superpotential shifts by
$$
W_{\rm eff}(S) \to W_{\rm eff}(S) + 2\pi i NS.
$$
 This multivaluedness then leads to the $NS \log S$ term in the
superpotential \vypot .  Note that this anomaly is closely related
to the measure of the path-integral, in particular that of the chiral
fermions.  Field theoretic aspects of this have been discussed in
\refs{\qfvy}.

In some sense minimizing the effective superpotential \vypot\ thus
turns a perturbative (one-loop) effect into the non-perturbative
condensate \gaugino.  This is a more familiar story in two-dimensional
supersymmetric linear sigma models such as the $\C\P^{N-1}$
model. There the 1-loop generated $N\Sigma\log\Sigma$ superpotential
also captures, after minimizing, the instantons of the two-dimensional
sigma model. In particular it gives rise to the quantum cohomology
ring
$$
\Sigma^N=e^{-t},
$$
where $t$ is the Kahler parameter of ${\bf CP}^{N-1}$ and $\Sigma$
represents the chiral field corresponding to the Kahler class. This
two-dimensional example of the quantum cohomology ring clearly
demonstrates that, even though we are in a massive vacuum, there is
still interesting holomorphic information to be computed---a point
that we want to stress also holds in the four-dimensional theory.

\subsec{An exact superpotential}

Now the question is how the form of the effective superpotential
$W_{\rm eff}(S)$ changes if we add the adjoint field $\F$ with the
tree-level potential \tree. The exact answer was given in \civ\ in
terms of a dual Calabi-Yau geometry given by the algebraic variety
\eqn\cy{
u^2+v^2 + y^2 + W'(x)^2 + f(x)=0,
}
where the polynomial $f(x)$ of degree $n-1$ is a deformation of the
singular geometry. Its coefficients parametrize the moduli of the CY
and thereby the gaugino condensates $S_i$ in the dual gauge
theory. This answer was checked against many non-trivial field
theoretic computations in \civ.

Furthermore in \cv\ it was explicitly verified that this answer can
also be derived from, and is in fact equivalent to, the Seiberg-Witten
solution of the undeformed $\cN=2$ $U(N)$ super-Yang-Mills theory
\refs{\sw,\klyt,\af}. The idea in \cv\ is to consider a very special
tree-level superpotential $W(\F)$ with degree $N+1$ for a $U(N)$ gauge
group. Thus $W'(\Phi)= \e \prod (\Phi -a_i)=0$ has $N$ roots $\Phi
=a_1,\ldots,a_N$.  If we place the $N$ eigenvalues of $\Phi$ at the
$N$ distinct values $a_i$ ({\it i.e.}\ the multiplicities are $N_i=1$
for all $i$), then as $\e \rightarrow 0$ we go back to a particular
point on the Coulomb branch of $\cN =2$ theory specified by the values
of the roots $a_i$.  It was shown in \cv\ that the $\cN=1$ theory has
certain non-trivial computable quantities that do not depend on $\e$
and therefore thus knows about the answer for $\cN=2$ which is obtained
as $\e \rightarrow 0$.  In this way the full Seiberg-Witten geometry
was derived from this $\cN=1$ deformation.

Then in \dv\ it was subsequently realized that this $\cN=1$ answer
has an elegant formulation directly in terms of the {\it perturbation}
theory of the original gauge theory.  It was found that the full
answer studied in \civ\ takes the form
\eqn\full{
W_{\rm eff}(S) = \sum_i \biggl[N_iS_i\log (S_i/\L_i^3)-2\pi i \tau S_i
+ N_i {\partial F_{\rm pert}(S) \over \partial S_i}\biggr], }
where $F_{\rm pert}(S)$ is a perturbative expansion in the $S_i$
$$
F_{\rm pert}(S) =\sum_{i_1,\ldots,i_n\geq 0}
c_{i_1,\ldots,i_n} S_{1}^{i_1}\cdots S_n^{i_n}.
$$
This perturbation series is obtained from the tree-level
superpotential $W(\F)$ by an expansion in terms of `fat' Feynman
diagrams around the classical vacua, only taking into account the {\it
constant} modes and the {\it planar} diagrams. Here every factor $S_k$
corresponds to a single loop (hole in the corresponding Riemann
surface) indexed by the critical point $a_k$. The coefficients
$c_{i_1,\ldots,i_n}$ are the planar amplitudes of the matrix model
with $i_k$ holes ending on the $a_k$ vacuum.  Such a series of terms
is not forbidden by non-renormalization effects because the explicit
breaking of $R$-symmetry by the tree-level superpotential, whose
coefficients have definite non-zero $R$-charge, gives a non-zero
effect in the background of a $S$ condensate. (We will argue for this
more precisely in the next section.)

Since only the planar diagrams contribute to $W_{\rm eff}$ the
$N$-dependence of the final answer is extremely simple. This
unexpected result, that follows straightforwardly from string theory,
is completely in accord with field theory arguments, as was shown in
\refs{\civ,\cv}.

The above prescription can be conveniently formulated in terms of the
following matrix model. Consider the saddle-point expansion of the
holomorphic integral over complex $N\times N$ matrices $\F$
\eqn\matrixint{
Z= \int d\F \cdot \exp\(- {1\over g_s} \Tr W(\F)\)  }
Since we only integrate over $d\F$ and not $d\F d\overline\F$, a
non-perturbative definition of this integral will require a specific
choice for the integration contour. But this choice will not influence
the perturbative expansion. For a generic position of the contour the
matrices $\F$ will be diagonalizable. The expansion parameter $g_s$
can be identified with the string coupling within a string theory
realization (since the tree-level superpotential is given by a disc
diagram) but within field theory it is simply a conveniently chosen
overall scale in front of the tree-level superpotential.

The usual large $N$ techniques tell us that around the saddle-point
where $N_i$ eigenvalues of $\F$ are at the $i$-th critical point, this
matrix integral has a consistent saddle-point approximation of the
form
$$
Z=\exp \sum_{g\geq 0} g_s^{2g-2} \cF_g(S)
$$
in the 't Hooft limit $N_i \gg 1$ and $g_s \ll 1$, while keeping finite the
individual 't Hooft couplings
$$
S_i = g_s N_i.
$$
We have one factor $S_i$ for each hole indexed by the index $i$. In
this large $N$ expansion of the free energy the term $\cF_g(S)$ is the
sum of all diagrams of genus $g$. In particular the leading
contribution $\cF_0(S)$ is given by the planar diagrams.

Furthermore, the contribution of the measure in the matrix model,
which is given by the volume of $U(N_1)\times\cdots\times U(N_n)$ (and
some gaussian measure which goes into the definition of the scales
$\L_i$) gives exactly rise to the $S_i\log S_i/\L_i^3$ term in the
superpotential\foot{If one would have included matter fields
transforming in the fundamental representation with mass matrix $m$,
then simply integrating out these gaussian variables in the matrix
model immediately gives the extra contribution $S \log\det (m/\L)$
that upon minimizing reproduces the Affleck-Dine-Seiberg
superpotential \ads.}. If we factor out these measure contributions,
we are left with
\eqn\pert{
F_{\rm pert}(S) =  \cF_{0,\rm pert}(S)
}
where $\cF_{0,\rm pert}(S)$ is given by the sum of all planar diagrams
contributing to the free energy of the matrix model.

To be completely specific, take for example the cubic superpotential
$$
W(\F)={\textstyle {1\over 2} m\F^2+{1\over 3} g\F^3},
$$
and consider the classical vacua where all eigenvalues sit at
$\F=0$. In that case the perturbative gauge group is still the full
$U(N)$. We are now predicting that the perturbative contributions to
the effective superpotential are obtained from the Feynman rules
derived from the action $W(\F)$, with propagator $1/m$ and three-point
vertex $g$. So the leading correction will take the form
$$
F_{\rm pert}(S) = ({1\over 6}+{1\over 2}){g^2\over m^3} S^3 + \cO(S^4),
$$
\ifig\twoloop{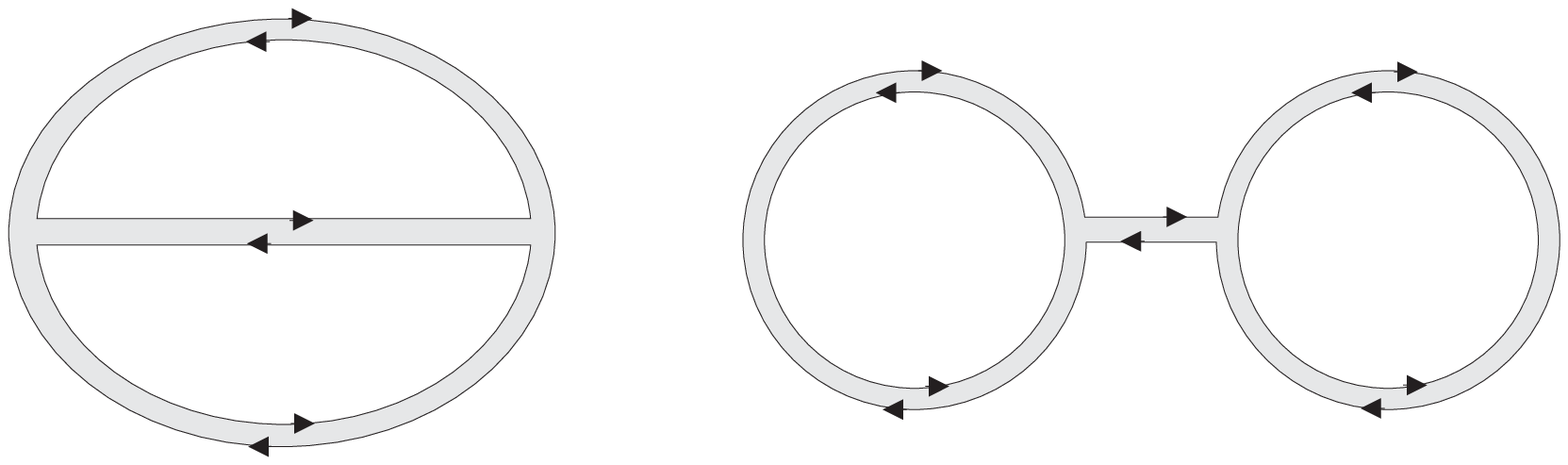}{100}{
The two planar two-loop diagrams, with combinatorial weight ${1\over
6}$ and ${1\over 2}$, that contribute to the order $S^3$ term in the
free energy.}
where the combinatorical weight in front is given by summing the two
planar two-loop diagrams of \twoloop. 
This indeed agrees with the
predictions in \civ .

Quite remarkably, this perturbative sum of Feynman diagrams will turn
into a non-perturbative sum over (fractional) instantons upon
minimizing with respect to the glueball superfields $S_i$. Even if one
cannot find an exact large $N$ solution, one could still work
order by order in the matrix loop expansion and translate this order
by order computation into an order by order computation for the
instanton sum.  In particular if we keep up to $k$-th term in the
$S_i$'s, after extremization we get corrections up to $k$-th fractional
instanton.  Moreover, summing all perturbative planar loops amounts to
an exact instanton sum.  Note that we have isolated a rare phenomenon,
a gauge theory quantity for which the planar limit is exact!

As we showed in \dv, in this specific example of a single adjoint
chiral matter field one can solve for the exact large $N$ result
directly using standard random matrices techniques. One diagonalizes
the matrix $\F$ and writes an effective action for the eigenvalues
$\l_1,\ldots,\l_N$
$$
S_{\rm eff}(\l)=\sum_I W(\l_I)- g_s \sum_{I<J}\log(\l_I-\l_J)^2.
$$ 
The derivative of this action for a given ``probe'' eigenvalue at a
position $x$ in the complex plane is then given by
$$
y(x)=W'(x) -g_s \sum_I {2\over x-\l_I},
$$
where the second term is a one-loop effect obtained by integrating out
the angular variables in $\Phi$. The one-form $ydx$ is directly related
to the eigenvalue density. In the large $N$ limit the variable $y$ can
be shown to satisfy an algebraic equation
\eqn\riem{
y^2=W'(x)^2+f_{n-1}(x)
}
that defines a hyperelliptic Riemann surface. Here $f_{n-1}(x)$ is a
polynomial of degree $n-1$ (the quantum deformation), whose $n$
coefficients can be used to parametrize the 't Hooft couplings $S_i$.
The solution of the matrix model then takes the form of a set of
period integrals (special geometry) of the meromorphic one-form
$ydx$
\eqn\periods{
S_i = {1\over 2\pi i} \oint_{A_i} ydx,\qquad
{\d \cF_0 \over \d S_i} = \int_{B_i} ydx,}
where $A_i$ and $B_i$ are canonically conjugated cycles on the Riemann
surface \riem. This Riemann surface and the one-form $ydx$ are the
reduction of the Calabi-Yau geometry \cy\ together with its
holomorphic three-form $\Omega$ to the variables $(x,y)$. This then
proves the equivalence between the perturbative computation of summing
planar diagrams and the dual Calabi-Yau geometry.

\subsec{The general conjectures}

We are now in a position to formulate our general conjecture. Consider
an $\cN=1$ supersymmetric gauge theory with gauge group $G$ a
classical Lie group, {\it i.e.}\ $G$ is a product of the gauge groups
of the type $U(N)$, $SO(N)$, or $Sp(N)$ (so no exceptional groups).
Moreover we consider a matter content compatible with an 't Hooft
Riemann surface expansion.  This is essentially equivalent to the
statement that the system has some kind of open string realization.
In particular one can assume we have a quiver type gauge theory
involving the product of series of $U(N_i)$ factors gauge groups with
some bi-fundamental matter content and some (single-trace) tree-level
superpotential terms. One might also possibly consider a $\Z_2$
orientifold of this system that gives rise in addition to $SO$ and
$Sp$ gauge groups which can be formulated directly at the level of the
quiver itself.  So for example spinor representations are not allowed,
and we can have up to second rank tensor products of fundamental
representation.  We also assume that we can add superpotential terms
to give mass to all the matter fields.

This system will have some tree-level superpotential for the chiral
matter multiplets $\F_a$
$$
\int d^2\!\theta\; W_{\rm tree}(\F_a).
$$
The classical vacua correspond to the critical points of $W_{\rm
tree}(\F_a)$. We will further assume that we are working with a vacuum
that is completely massive, possibly up to some pure $\cN=1$
Yang-Mills theory with group $G=\times_i G_i$ (each factor of which is
associated to an integer $N_i$).

We want to determine the effective superpotential
$$
\int d^2\!\theta\; W_{\rm eff}(S_i)
$$
in terms of the gaugino bilinear fields $S_i$ of the gauge groups
$G_i$. Our claim is now the following: the effective superpotential is
always of the form
$$
W_{\rm eff}(S_i)=\sum_i \({\widehat N}_i S_i \log (S_i/\L_i^3)-2\pi
i\tau_i S_i\) + W_{\rm pert}(S_i)
$$
with $\tau_i$ the bare couplings. The leading terms is the usual one
loop matching of scales.  For the $U(N_i)$ gauge factors the integers
$\widehat N_i$ are given by the rank ${\widehat N}_i=N_i$; for the
case of $SO(N_i)$ and $Sp(N_i)$ they are given by ${\widehat
N}_i=N_i\mp 2$.  The perturbative contributions $W_{\rm pert}(S_i)$
come from two sources: 't Hooft diagrams with the topology of $S^2$ or
$\R\P^2$ (the latter arises only when we have the orientifold
operation and the resulting unoriented Riemann surfaces).

We will argue that these perturbative contributions are computed by
the associated matrix model with action given by $W_{\rm tree}(\F_a)$.
Consider the saddle-point expansion of this matrix theory
corresponding to the vacuum with unbroken gauge group $\times_i G_i$.
Let $\cF_0(S_i)$ and $\cG_0(S_i)$ denote the contributions to the free
energy of the matrix model of diagrams with the topology of $S^2$ and
$\R\P^2$ respectively. Here the $S_i$ dependence of the diagrams
captures the number of 't Hooft loops ending on the group $G_i$ and
thus can be identified with the coupling $g_s {\widehat N}_i$. Then we
claim that
$$
W_{\rm pert}(S_i)={\widehat N}_i {\partial \cF_0(S_i)\over \partial
S_i}+\cG_0(S_i).
$$
(The one loop $S\log S$ piece can also absorbed into the above if we
include the correct measure for the matrix theory.)

Moreover for the abelian $U(1)\subset G_i$ factors, that remain after
the non-abelian gauge factors develops a mass gap, one can find the
matrix of coupling constants $\tau_{ij}$. It is also given in terms of
the planar diagrams as
$$
\tau_{ij}={\partial^2 F_0\over \partial S_i \partial S_j}.
$$
One can extend this conjecture to include certain gravitational
corrections as will be discussed below.

Here we have formulated our conjecture in the context of $\cN =1$
theories with no massless moduli.  However this correspondence also
holds in the cases where we have massless moduli.  To see this, note
that we can always deform the superpotential and give mass to the
massless moduli, reducing to the case considered in this conjecture,
and then in the end removing the deformation. This idea will be
important for us in this paper in the context of $\cN=2$ and $\cN=4$
supersymmetric gauge theories, which will have massless moduli. More
generally, in the case of moduli there is still interesting
holomorphic data to be computed, like the chiral ring of operators,
and we claim that these data too should be captured by planar
diagrams.

\subsec{Calabi-Yau geometry as large $N$ master field}

It is not necessary to solve for the exact large $N$ limit of the
matrix model if one is content to work order by order in the instanton
expansion. However, in case a large $N$ solution is possible we are
conjecturing that the master field configuration that dominates in the
large $N$ limit, will take the form of the special geometry associated
to a (non-compact) Calabi-Yau three-fold. In the previous example, and
also in the case of the quiver matrix models considered in
\dvii, this three-fold is of the special form
$$
u^2+v^2+F(x,y)=0,
$$
and is thus essentially given by the complex curve $F(x,y)=0$. But in
more complicated cases we expect this geometry to be a truly
three-dimensional complex manifold. For example, as a first step in
this direction, in \ckv\ a quiver gauge theory is described with two
adjoints and a tree-level superpotential
$$
W_{\rm tree} = \Tr\(\F_1\F_2^2 + \F_1^n\).
$$
It is further shown that the exact effective superpotential is
computed in terms of an associated dual CY geometry (``Laufer's
geometry'') that is of the more involved form
$$
u^2 + F(v,x,y)=0,
$$
where $F(v,x,y)=0$ defines a complex surface. It would be very
interesting to derive this geometry directly from the matrix model, in
particular to find the role of the three-form $\Omega$, that in this
solution will reduce to a meromorphic two-form on the
complex surface.

\newsec{Derivation of our conjecture}

We will now argue that for any $\cN=1$ gauge system that can be
embedded in string theory, for example by engineering a suitable
configuration of D-branes in a Calabi-Yau background in type II
theory, we can give a proof of our main conjecture. As we will show
below, that argument can be completely reduced to the $\a'\to 0$ field
theory limit---no stringy effects play any role, otherwise than that
the gauge should allow for a diagrammatic large $N$ expansion. We are
confident therefore that this argument can ultimately be given
completely in field theoretic terms.

One case where this strategy has been worked out successfully was the
case of the conformally invariant orbifolds of $\cN=4$ gauge theory
introduced in \refs{\ks,\lnv}. The vanishing of the beta function to
leading order in large $N$ was later shown to be equivalent to that of
the original $\cN=4$ theory, but the arguments for that remarkable
result relied on string theory perturbation techniques \bkv. However
as noted in \bkv\ one should have expected these results to also be
derivable directly in field theory by taking the $\a'\to 0$ limit of
string perturbation theory and in fact such a direct proof was later
given entirely within field theory \bj.  However the insight of string
theory was crucial for coming up with such an argument.  The same is
true in the case we are studying.

\subsec{F-terms in string theory}

As was first discovered in \refs{\bcov,\agnt}, F-terms for a
four-dimensional supersymmetric theory correspond within type II
superstring theory to a particular class of amplitudes that are
exactly computed by topological string theory. The topological string
is entirely formulated in terms of the internal CY space, and so is in
some sense localized to constant modes in space-time. For type I open
strings there is also such a correspondence between F-terms and open
topological strings that was found in \refs{\bcov,\S8.2} where it was
further shown that the superpotential for glueball fields receives
contributions only from planar diagrams.  The possibility of applying
this powerful technique to the dynamics of four-dimensional
supersymmetric gauge theories was already noted. The arguments in
\bcov\ easily generalizes from type I strings to arbitrary D-branes
(whose relevance was not yet appreciated around that time).
The derivation of these results is particularly simple
in the Berkovits formalism, generalizing the same
computation in the context of closed topological strings
done in \berva.

\ifig\cuts{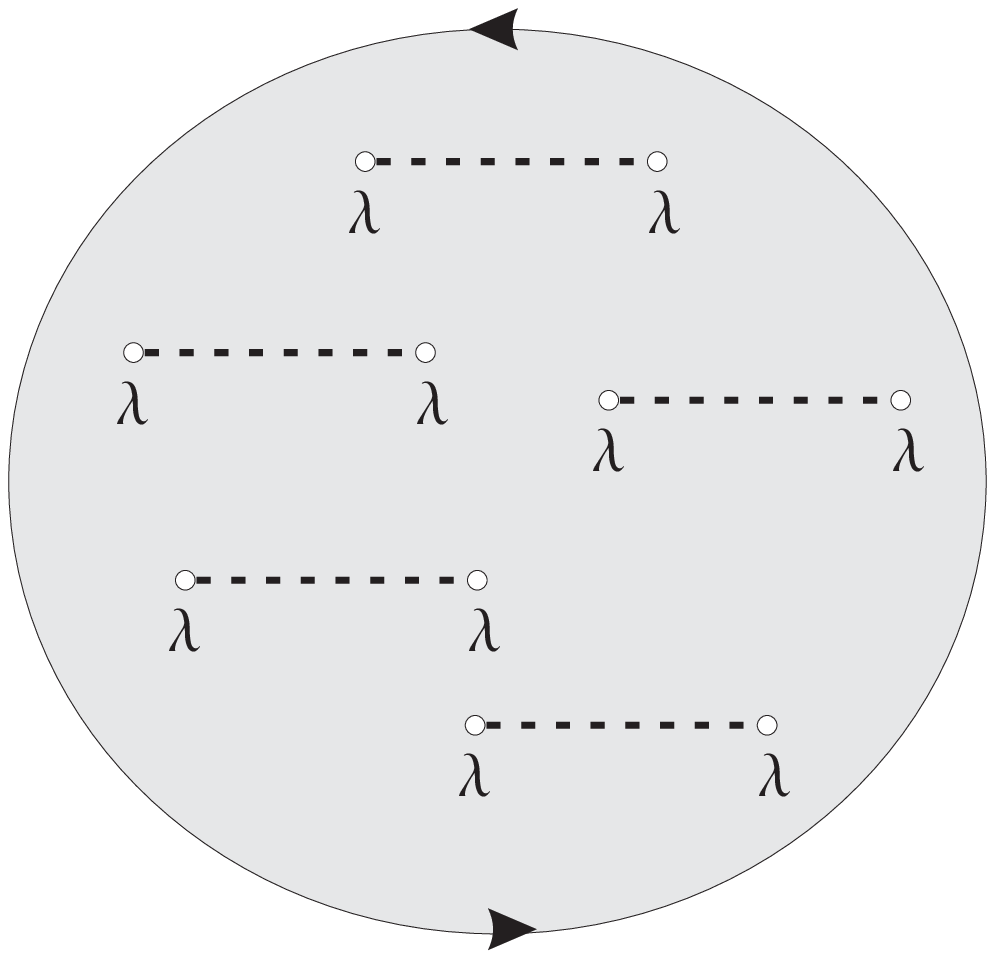}{45}{A planar open string diagram contributing to the
superpotential as represented in topological string
theory. Zero-momentum vertex operators for the gaugino $\l$ are
inserted at the end points of the cuts that represent the open string
loops.}

Let us briefly review why in the string context the superpotential is
completely captured by the planar diagrams. There is a simple
geometric argument given in \bcov. One realizes the open string
diagram with a very particular worldsheet metric, namely by a flat
metric with cuts in it---very much as in light-cone gauge.  This
metric has curvature singularities at the two end points of each cut,
where the deficit angle is $2\pi$ instead of $\pi$. These curvature
insertions correspond exactly to insertions of the gaugino vertex
operator {\it at zero momentum}, as sketched in \cuts.

Since all insertions are at zero-momentum, the topological string
amplitude is computing a non-derivative term in the effective
action---the effective superpotential. The topological twisting
produced by the insertions of these vertex operators induces a
complete cancellation of contributions of the four-dimensional bosonic
and fermionic fields. The twist gives them the same worldsheet spins
and therefore they cancel completely in the worldsheet
path-integral---including the bosonic zero-modes, so there is no
four-dimensional momentum flowing through the loops.

The combination of two of these insertions at each end point together
with the sum over Chan-Patton indices gives exactly the required
factor of a single $\Tr\,\l_{i}^2 = S_i$ insertion for each hole in
the string worldsheet. Zero-mode analysis gives that only for planar
diagrams such an amplitude with only gaugino insertions can be
non-vanishing.  Moreover, each hole gets two gaugino fields except for
one boundary hole. This is necessary to have two unabsorbed fermionic
zero-modes, since we want a space-time term of the form $\int\!
d^2\theta \, W_{\rm eff}(S)$.

\ifig\weff{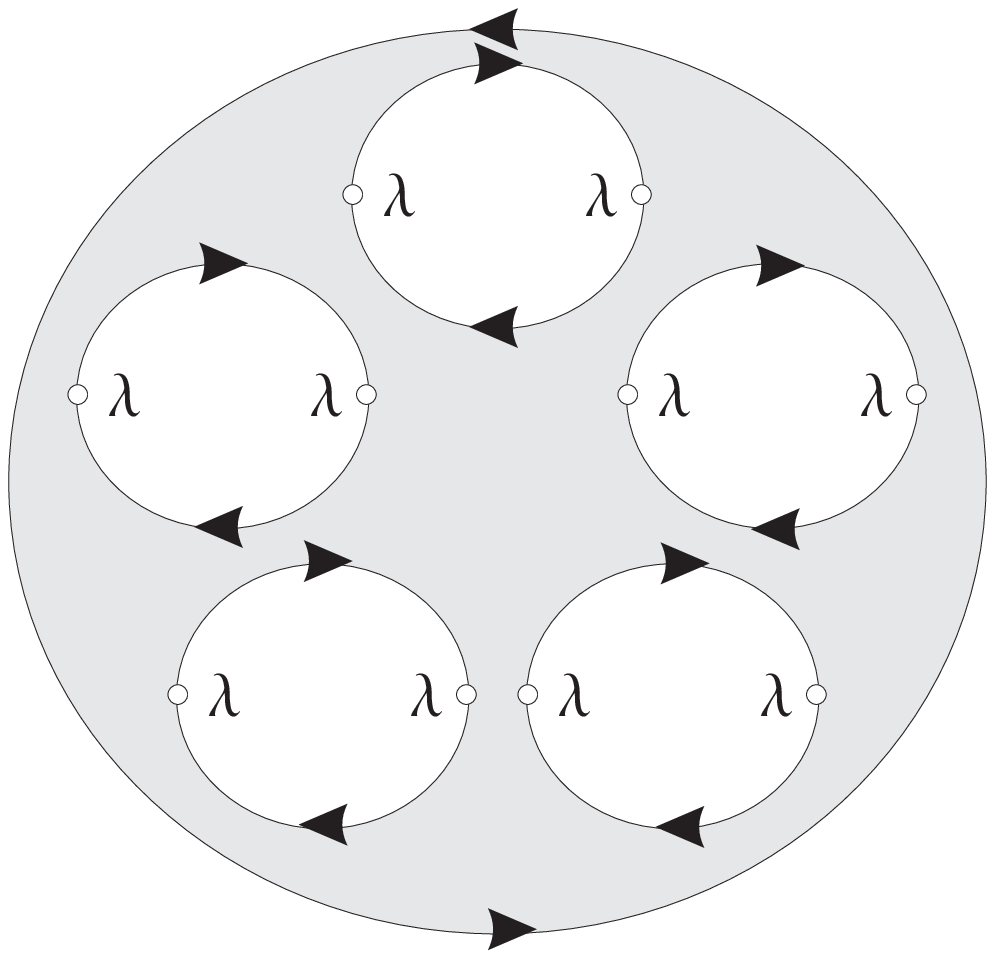}{45}{
A typical worldsheet diagram that contributes to the
superpotential $W_{\rm eff}(S)$. The outer index loop has no
$S=\Tr\,\l^2$ insertion.}

So from this one concludes that the perturbative piece to the
superpotential is given exactly by a series of the form (we work for
convenience with one set of Chan-Paton indices of rank $N$)
$$
W_{\rm pert}(S) = \sum_{h>0} Nh \cF_{0,h} S^{h-1}=N{\d \cF_0(S)
\over \d S},
$$
where $\cF_{0,h}$ is the open topological string amplitude with $h$
holes. The factor $Nh$ arises from choosing a hole for which no
gaugino $\l_\a $ is inserted and the factor $N$ represents the
corresponding contribution from the Chan-Paton factor \vaug, see
\weff.  This combinatoric factor gives rise to the $N\partial/\partial
S$ structure seen in the above formula.

Zero-modes analysis allows for another configuration. One can
also distribute one of the gaugino fields from the inner holes
to the outer hole.  In other words, two holes will have a single
gluino vertex operator and the rest of the holes will each have two.
These diagrams will compute the exact coupling constant of the
low-energy $U(1)$ gauge fields.  If we put a single gluino $\l_\a$
(instead of two) at two of the loops, the Chan-Paton factor does not
kill it and this gives rise to a correction for the coupling constant
of the corresponding $U(1)$, leading to the formula given before:
$$
\tau_{ij}={\d^2 \cF_0(S) \over \partial S_i\partial S_j}.
$$
\ifig\tauij{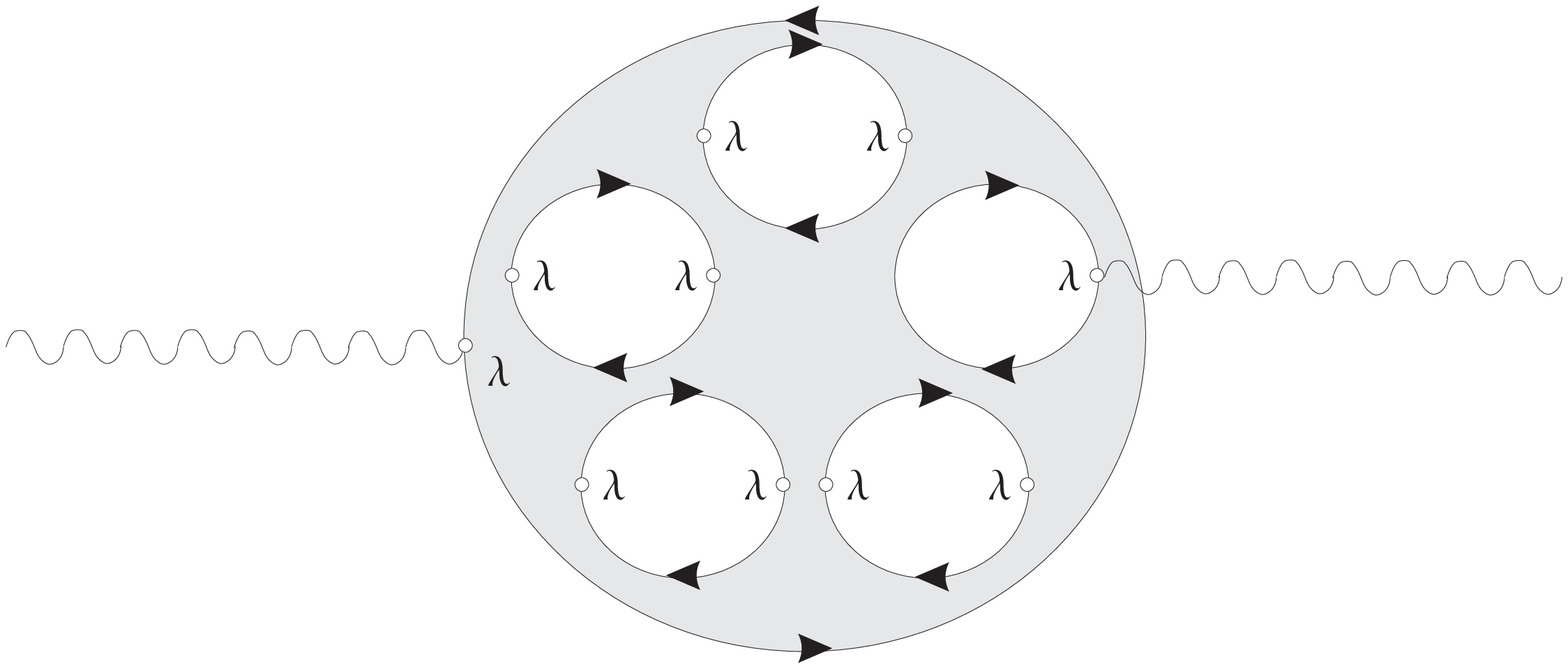}{95}{
A planar worldsheet diagram that contributes to the coupling
$\tau_{ij}(S)$ of the $U(1)$ gauge fields. There are two holes with
only a $\Tr\,\l$ insertion; all other holes carry a $\Tr\,\l^2$
insertion.}  The typical diagram contributing to this gauge coupling
is sketched in
\tauij.

Higher genus surfaces require bulk, {\it i.e.}\ gravitational, vertex
operators (such as a graviphoton vertex) and thus do not contribute to
the glueball superpotential or the coupling constant of the $U(1)$'s.

Within type IIB string theory there are no non-trivial worldsheet
instantons or other $\a'$-effects that correct these result, and one
can therefore take the $\a'\to 0$ (infinite tension) limit where all
excited string modes decouple and the string diagrams reduce to
ordinary ``fat'' Feynman diagrams.  This gives the proof of the above
conjecture for all the cases that can be engineered by type IIB string
theory.  However, it also looks feasible to reproduce the powerful
stringy zero-mode arguments directly in four-dimensional field theory,
thus generalizing the proof of our conjecture to cases beyond those
realized in string theory. Let us sketch a possible approach to such a
field theory argument. 

\subsec{Sketch of a field theoretic argument}

Within field theory one basically considers what kind of path-integral
configurations can contribute to superpotential terms.  These come
from configurations which preserve two of the four supercharges
(giving rise to a $d^2\theta$ integral).  In this way the
four-dimensional supersymmetric gauge theory path-integral
automatically localizes to configurations where the $\Phi$ field is
constant (by the usual argument that the anti-chiral supersymmetry
variations of the fermionic fields gives total derivative terms
$\partial_\mu\Phi$, which must thus be zero).  Thus the computation
reduces directly to an action of the form $\int
\! d^2\theta\; W_{\rm tree}(\Phi)$.  Let us for simplicity assume that
$$
W_{\rm tree}(\F) = {1\over 2} m\F^2 + {1\over 3}g \F^3.
$$
To find the perturbative corrections to the Veneziano-Yancielowicz
superpotential one could now argue as follows. Since by assumption we
are in a massive vacuum, and since we are computing a $d^2\theta$
term, we can work entirely in the zero-momentum sector. This allows
one to completely discard the D-term and only work with the chiral
F-term. From this point of view the propagator for the chiral
superfield $\F(\theta)$ is coming directly from the quadratic part in
the superpotential
$$
\int\! d^2\theta\; \half m\F^2.
$$ 
This is unconventional, since usually the kinetic terms come from
integrating out the auxilary fields using the D-term and couple $\F$
to $\overline\F$. Similarly the interacting vertices are coming from
the higher order terms in $W_{\rm tree}(\F)$. Working in a
supercoordinate basis, we have the propagator
$$
\<\F(\theta)\F(\theta')\> = {1\over m} (\theta-\theta')^2
$$
with $\theta^2 \equiv \e_{\a\b}\theta^\a\theta^\b$. Each vertex gives
a factor $g \int\! d^2\theta$ which removes two $\theta$'s. A simple
counting of factors of $\theta$ for a diagram with $V$ vertices and
$P$ propagators gives a net factor of $n=2(P-V)$ fermionic
zero-modes. For a diagram with $h$ index loops and with topological
genus $g$ we have $n=2(h+2g-2)$.

We now claim that in a gaugino condensate background these extra
$\theta$'s can be absorbed, but only for planar diagrams. If we
naively couple to a background gaugino field $\l_\a(\theta)$ we have
additional vertices of the form $\l_\a(\theta) \d/\d\theta_\a$ and
these can absorb extra $\theta$'s.  The insertion of the single-trace
glueball field $S(\theta)=\Tr\,\l_\a^2$ corresponds to two of these
$\l_\a$ vertices, and by the index structure such an insertion is
associated to a single index loop. So in this way we can absorb at
most $2h$ zero-modes. Therefore only in the case of planar diagrams
can insertions of $S(\theta)$'s completely remove the fermionic
zero-modes. If one inserts in a planar diagram an $S(\theta)$ for all
loops except one, one removes exactly all the remaining $\theta$'s
from the propagators, leaving a final overall $\int \!  d^2\theta$
integral, giving the required structure for a term of the form $\int\!
d^2\theta\,W_{\rm eff}(S)$.

It would be interesting to see whether this argument can be made into
a more precise derivation; this is currently under investigation.  As
for the open string diagrams with the topology of $\R\P^2$ that arise
in the context of orientifolds, the arguments of \bcov\ can be
extended as discussed in \refs{\sinhv,\aahv} leading to the
contribution $\cG_0(S)$ given above.

\subsec{Gravitational couplings}

One would wonder what the non-planar diagrams are good for?  Again
string theory provides the answer and thus extends our general
conjecture to include certain gravitational correction. To obtain the
topological amplitudes from ordinary superstrings in the case of
non-planar diagrams we need to insert spin operators in the bulk of
the worldsheet, which correspond to closed string or gravitational
vertex operators.  A completely similar analysis gives that for a
worldsheet of genus $g>0$, we obtain a term in the four-dimensional
effective action of the form \vaug :
$$
\int \!d^4\!x \!\int\! d^2\!\theta \sum_{h\geq 1}N h\, \cF_{g,h}
S^{h-1} (\cW^{2})^g +
\int \!d^4\!x \!\int\! d^4\!\theta
\sum_{h\geq 1}
\cF_{g,h} S^{h} {\cW^2}^g.
$$
Here $\cF_{g,h}$ indicates a contribution of a worldsheet with $g$
handles and $h$ holes; $\cW_{\alpha \beta}$ is the gravitational Weyl
multiplet in $\cN=2$ supergravity. The lowest component of this chiral
field is the self-dual part graviphoton field strength $F_+$, the
highest component is the self-dual part of the Riemann tensor $R_+$.

Expanding the superspace integral in components for genus $g=1$ we get
a term that measures the response of coupling the $\cN=1$ gauge theory
to a non-trivial four-dimensional gravitational background by the
effective action
\eqn\onlo{
\int \! d^4x\, \cF_1(S) \Tr\,R_+^2.
}
\ifig\genusone{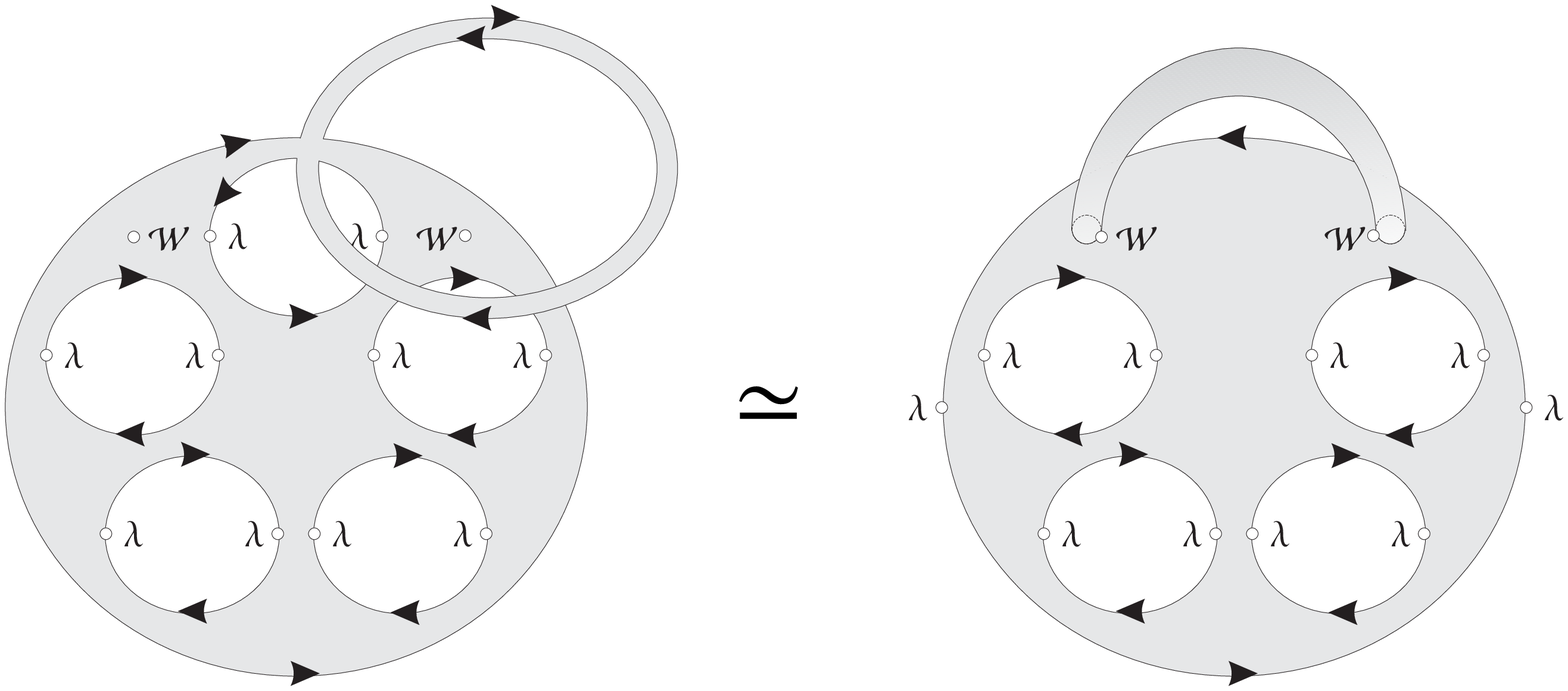}{110}{A genus one worldsheet diagram that 
contributes to the gravitational coupling $\cF_1(S)\cW^2$. Note that
there are now two bulk insertions of graviton vertex operators $\cW$,
corresponding to the insertion of the extra handle compared to the
planar diagrams.} 
A typical worldsheet diagram contributing to this
term is given in \genusone.

More precisely, including the normalizations, we get a term
$$
{1\over 2}\cF_1(S) (\chi -{3\over 2}\sigma),
$$
where $\chi$ and $\sigma$ denote the Euler characteristic and
signature of the four manifold respectively.  Here the coefficient
$\cF_1(S)$ is given as a sum over all genus one diagrams with an
arbitrary number of holes
$$
\cF_1(S) = \sum_{h\geq 0} \cF_{1,h}  S^{h}
.
$$
In particular we find from the planar limit the expectation value of
the gaugino condensates $S_i$ and plug it into this formula to find
the gravitational corrections.  It was suggested in \vaug\ that the
higher genus corrections for $g>1$ in addition measure the response of
the gauge theory to a non-commutativity in spacetime represented by
the (self-dual) graviphoton field strength $F_+$. This idea has found
an interesting application in the recent computation of the instanton
effects in the $\cN=2$ Yang-Mills theory \nekrasov.

If there is a dual Calabi-Yau geometry with moduli $S_i$ that captures
the large $N$ limit, $\cF_1(S)$ is given by the analytic torsion on
that manifold \bcov.  If the geometry essentially reduces to a Riemann
surface $\Sigma$ given by an algebraic curve
$$
F(x,y)=0,
$$
as was the case for the perturbed $\cN=2$ theories in \dv\ and the
quiver cases of \dvii, there are general arguments (basically from the
fact that the full partition function is given by a tau-function of an
integrable hierarchy) that $\cF_1$ is given by the logarithm of the
determinant of a chiral boson on the Riemann surface $\Sigma$ (see
{\it e.g.}\
\kostov)
\eqn\onch{\cF_1(S) = - \half \log \det \dbar_\Sigma.}
We will make use of this result when we consider the $\cN=4$
theory later in this paper.

\newsec{Perturbative derivation of dualities}

In this paper we have proposed a general method to compute IR effects
including exact instanton contributions to F-terms for a large class
of $\cN=1$ theories.  It is natural to ask if this perturbative
perspective carries non-perturbative information such as
$S$-dualities.  Indeed we will point out how Seiberg-like dualities,
are derivable from this perspective.  It is also natural to ask if we
can derive non-trivial exact results for theories with more
supersymmetry, such as $\cN=2$ and $\cN=4 $ theories.  The answer to
this also turns out to be positive.

For example, consider the $\cN=4$ theories deformed by mass terms for
the adjoint fields.  The deformation is known as the $\cN=1^*$ theory
and was introduced in \vw\ to analyze topologically twisted $\cN=4 $
theories on 4-manifolds.  Here we will be able to make contact with
the results in \vw\ from our direct perturbative analysis.  Moreover
the value of the superpotential has been studied for these theories in
\refs{\dorey} and we will derive this result summing the
planar diagrams below.  Quite remarkably we will derive in this way,
using only perturbative techniques, the modular properties of the
superpotential.

We can also ask whether we can recover the exact results of
Seiberg-Witten geometry for $\cN=2$ theories. This also turns out to
be possible.  We thus see that our perturbative methods are not only
rather powerful, but we have also a derivation of these non-trivial
conjectured dualities from first principles.

In this section we first briefly indicate how Seiberg-like dualities
can be derived in this setup.  We next discuss how the exact $\cN=2$
results are recovered.  Then we turn to the main topic of this
section, the $\cN=4$ theories and their deformations.

\subsec{Seiberg-like dualities}

Consider for definiteness the A-D-E $U(N_i)$ quiver theories with
bi-fundamental matter and with adjoint matter with some superpotential
for each adjoint field.  Then, as shown in \cfikv\ Seiberg-like
dualities correspond to Weyl reflections of the nodes of the Dynkin
diagram.  Moreover, it was shown in \cfikv\ that these dualities are
manifest in the superpotential for the glueball fields.  Basically,
the glueball field superpotential is insensitive to Weyl reflections,
up to appropriate field redefinition.  As we have discussed here and
in \dvii , the corresponding A-D-E matrix models at the planar limit
compute the glueball superfield superpotential, and the large $N$
solution reproduces the geometry of \cfikv.  It follows that
Seiberg-like dualities are visible in the perturbative gauge theory
analysis of the glueball superpotential which reduce to planar
diagrams of the A-D-E matrix models.

\subsec{The $\cN=2$ Seiberg-Witten solution}

For simplicity let us consider the case of pure $\cN=2$ supersymmetric
$U(N)$ Yang-Mills, though the statement can be easily generalized to
quiver type theories.  Consider deforming the theory by a
superpotential $W(\F)$ which is a polynomial of degree $N+1$ in the
adjoint field $\F$.  In particular we have
$$
W'(\F)=\e P_N(\F) =\e \prod_{i=1}^N (\Phi -a_i).
$$
Choosing the branch where each eigenvalue of $\Phi$ is equal to one of
the roots $a_i$, we can freeze to an arbitrary point on the Coulomb
branch given by $a_i$.  The $U(N)$ gauge group breaks to $U(1)^N$ for
this branch.  Then, as shown in \cv, extremization of the glueball
superpotential, which as we have argued is perturbatively computable
by summing planar diagrams, leads to the special geometry on the curve
$$y^2=P_N(x)^2-\Lambda^{2N}$$
which is in exact agreement with the Seiberg-Witten curve for this
case \refs{\sw,\klyt,\af}. We stress once more that in our approach
this geometry appears directly out of the planar limit of the
perturbative diagrams---no non-perturbative duality is
invoked. Moreover the coupling constant for the $U(1)^N$ theory is
given by
$$
\tau_{ij}={\partial^2 \cF_0 \over \partial S_i\partial S_j}
$$
which just gives the period matrix for this geometry.  Note in
particular that this result is independent of $\e$ and taking
$\e\rightarrow 0$ reduces the problem to that of unbroken $\cN=2$
gauge theory, in agreement with the predictions of the SW geometry.

Similarly one could ask about the computation of the BPS masses
$a,a_D$.  In this case it was found in \cv\ that there is a reduced
1-form $h=P_N'(x)dx/y$ (corresponding to the $H$-flux in the
underlying Calabi-Yau geometry) which plays the role of the smeared
density of the eigenvalues of the $\Phi$ field (which is the source of
$H$-flux).  In particular it has normalized periods $\oint_{A_i} h=1$.
Thus the average value of $\Phi$ which at the $i$-th vacuum is
classically given by $a_i$ gets replaced by the quantum expression
$$
a_i^{quantum}=\langle x\rangle =\oint_{A_i} x\, h
$$
and as shown in \cv\ this exactly corresponds to the periods of the SW
differential.

\subsec{The $\cN=4$ theory and $S$-duality}

A particular interesting case to apply our general philosophy is the
$\cN=4$ super-Yang-Mills theory that is well-known to have a
strong-weak coupling Montonen-Olive $S$-duality \mo.  Some highly
non-trivial aspects of this duality were checked in \vw. By now this
$S$-duality is generally accepted and it has become the foundation on
which many other dualities, both in field theory and string theory,
rest. Can non-trivial consequences of this non-perturbative duality be
seen using our perturbative large $N$ techniques? The answer is,
surprisingly, yes.

The $\cN=4$ $U(N)$ gauge theory can be considered as a $\cN=1$ gauge
theory with three adjoint chiral superfields $\F_1,\F_2,\F_3$ with a
tree-level superpotential $\Tr\(\F_1[\F_2,\F_3]\)$.  We will first
consider breaking this theory softly down to $\cN=1$ by introducing
masses for all the adjoints, {\it i.e.}\ working with the tree-level
superpotential
$$
W_{\rm tree}(\F)= \Tr\(\F_1[\F_2,\F_3] + \sum_i m \F_i^2\).
$$
This is the so-called $\cN=1^*$ theory.  This deformation was
introduced in \vw\ to compute the topologically twisted $\cN=4$ theory
on some manifolds including $K3$.  In the context of AdS/CFT dualities
this deformation was studied in \ps.

According to our philosophy the effective superpotential of this model
should be captured by the planar diagrams of the following
three-matrix model, expanded around the relevant vacuum
\eqn\threemm{
\int d\F_1 d\F_2 d\F_3 \; \exp
\Tr\(\F_1[\F_2,\F_3] + \sum_{i=1}^3 m \F_i^2\).
}
Remarkably this model has been studied---and solved---in \kkn\
(building, as we understand, on earlier work in \hoppe) where it was
considered in relation with the matrix models of M-theory. We will
review the method of that paper, and present our own interpretation of
the main result. In the process we will present the solution in a
language that is more closely tailored to our needs.

For simplicity we will consider only the classical vacua given by the
trivial configuration
$$
\F_i=0.
$$
Here we have perturbatively still the full $U(N)$ gauge symmetry, but
quantum effects will produce a mass gap and $N$ confining vacua. This
choice of perturbative vacuum correspond to the usual small
fluctuations saddle-point approximation of \threemm.  The main fact
that allows one to solve the matrix model is that, with this choice of
vacuum, one can consider the action to be quadratic in the matrices
$\F_2,\F_3$ and integrate them out. Indeed going to a basis $\F_\pm =
\F_2 \pm i
\F_3,$ we can write the action as
$$
W_{\rm tree}(\F) = \Tr\(i\F_+[\F_1,\F_-]+m\F_+\F_- + m \F_1^2\).
$$
Therefore integrating out $\F_\pm$ as in a gaussian integral gives an
extra determinant in the effective action for the remaining adjoint
$\F=\F_1$. The matrix integral reduces in this way to the following
holomorphic one-matrix model
$$
\int d\F {e^{m\Tr\,\F^2} \over \det\([\F,-]+im\)}.
$$
(Clearly the derivation this far would have worked for any potential
$W(\F)$, not necessarily quadratic, for the first adjoint. We leave
this generalization for further study.)  The induced measure looks
dangerously complicated but we can still go to an eigenvalue basis
where it reduces to
\eqn\lll{
\int\prod_I d\l_I \prod_{I<J} {\(\l_I-\l_J\)^2 \over
\(\l_I-\l_J +im\)\(\l_I-\l_J-im\)} \exp {\sum_I m \l_I^2}.
}

This result has an obvious interpretation as a gas of $N$ eigenvalues
$\l_I$ of charge $+2$ in a potential $W(x)=x^2$, interacting not only
through the usual Coulomb potential with each other (as expressed by
the numerator), but also with two charge $-1$ mirror images shifted by
$\pm im$ in the complex plane (as indicated by the denominator). Note
however that there are no mutual interaction among these two mirror
images.

As a side remark we point out that, if we put the mass deformation for
$\F_2$ and $\F_3$ to zero, which in field theory language corresponds
to a flow to a non-trivial conformal fixed point \kpw, the total
measures in \lll\ becomes completely trivial. This is a reflection of
a more general fact that four-dimensional CFT's correspond to net
zero-charge Coulomb gas models.

\subsec{The large $N$ solution}

After rescaling the eigenvalues $\l_I \to m \l_I$ and writing
$g_s=1/m^3$ (a complex parameter) the equation of motion reads
$$
2\l_I = g_s \sum_{J \not= I} \Bigl\{ {2\over \l_I -\l_J}-{1\over
\l_I-\l_J+i} -{1\over \l_I-\l_J-i} \Bigr\}.
$$
The only parameter in the large $N$ limit will be the 't Hooft
coupling
$$
S=g_sN.
$$
The solution proceeds, as always, through study of the 
resolvent
$$
\w(x) = {1\over N} \sum_I {1\over x - \l_I}.
$$
The general dynamics of large $N$ matrix models tells us that in this
limit the eigenvalues will spread out from their classical locus
$\l_I=0$ at the minimum of the potential well, into (in this case) a
single cut $(-a,a)$ along the real case. The size $a$ of the cut is
determined by the 't Hooft coupling, and is the only parameter in the
solution.  The continuous eigenvalue density $\rho(x)={1\over N}
\sum_I \delta(x-\l_I)$ is given by the jump in the resolvent
$$
\rho(x)=-{1\over 2\pi i}\Bigl[\w(x+i\e)-\w(x-i\e)\Bigr].
$$
The force $f(x)$ on a probe eigenvalue $\l_I=x$ in the complex plane
takes the form
$$
f(x)=2x - S \Bigl[2\w(x)-\w(x+i)-\w(x-i)\Bigr].
$$
It is zero by the equation of motion along the cut. Following \kkn\ we
further introduce the function
\eqn\yyy{
G(x)=x^2 + i S\Bigl[\w(x+{i\over 2})-\w(x-{i\over 2})\Bigr], }
that is related to the force $f(x)$ through the relation
\eqn\force{
f(x)=-i\Bigl[G(x+{i\over 2})-G(x-{i\over 2})\Bigr].
}
Note that on the cut $x\in(-a,a)$ we have
$$
G(x+{i\over 2})=G(x-{i\over 2}).
$$
One furthermore has the relations $G(-x)=G(x)$ and (for real $S$)
$G(\overline x)=\overline{G(x)}$, so that $G(x)$ is completely
determined by its values in one quadrant of the $x$-plane.

\subsec{The elliptic curve}

The above relations actually imply somewhat surprisingly that $G(x)$
is defined on an elliptic curve \kkn. It is this elliptic curve, that
emerges as the master field in the large $N$ limit, that in the end
will imply the modularity of the superpotential and the $S$-duality of
the underlying $\cN=4$ theory.

\ifig\elliptic{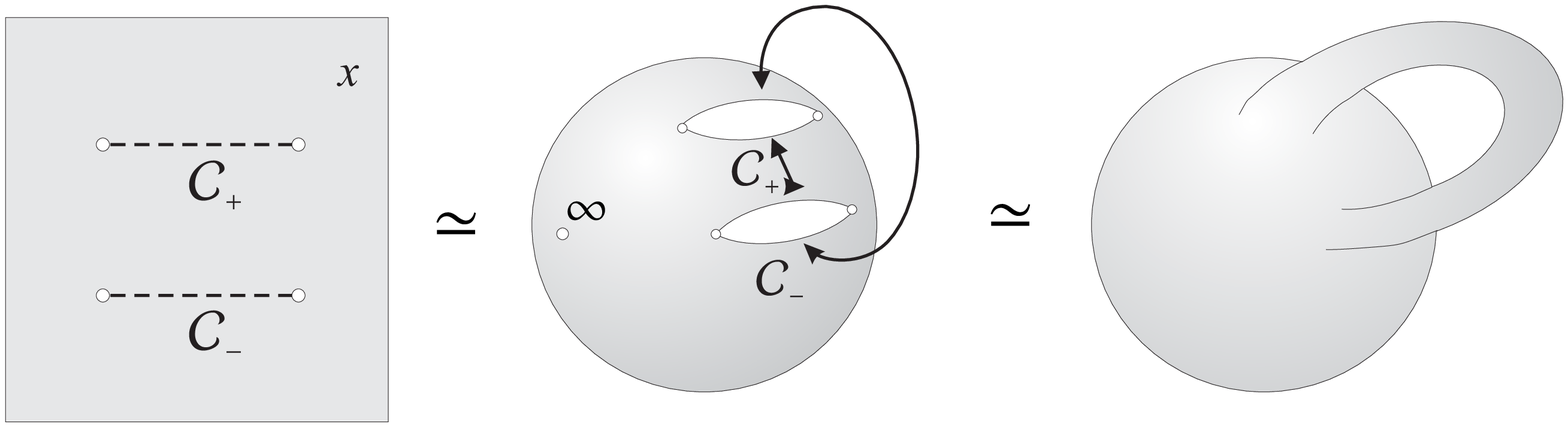}{100}{
The emergence of the elliptic curve from the geometry of
the matrix model.}

This elliptic curve is determined by the behavior of $G(x)$ on the
$x$-plane. The function has discontinuities along the two cuts
$$
\cC_+= (-a+{i\over2},a+{i\over2}),\qquad \CC_-=(-a+{i\over2},a+{i\over2}).
$$
These cuts are simply the two reflections of the original cut on which
the eigenvalues condense. We can now think of the complex $x$-plane,
compactified by adding the point at infinity, as a two-torus, by
cutting the two cuts $\cC_+$ and $\cC_-$ open and then gluing them
together. In this gluing we identify the upper-half of $\cC_+$ with
the lower-half of $\cC_-$ and {\it vice versa}. So, pictorially we
make two incisions in the Riemann sphere and glue in a handle
connecting the cuts to make a genus one Riemann surface, see 
\elliptic.

To work with this elliptic curve in a more traditional representation
we follow \kkn\ and observe that the function $t=G(x)$ gives the
$x$-plane as a branched double cover of the $t$-plane. More precisely
$G(x)$ maps the upper quadrant in the $x$-plane (folded around half of
the cut $\cC_+$) to the upper-half-plane in $t$.
This can be seen by inverting $G(x)=t$ and write
$$
x= G\inv(t)= A\int_{x_1}^t {(t-x_3) dt\over \sqrt{(t-x_1)(t-x_2)(t-x_4)}}
$$
for appropriate constants $A,x_1,x_2,x_3,x_4$. Here the branch points
$t=x_1,x_2,x_4$ get mapped to $x=0,{i\over2}-i\e, {i\over 2}+i\e$, and
the regular point $t=x_3$ gets mapped to $x=a+{i\over 2}$, the end
point of the eigenvalue cut. All the constants are in principle
completely fixed by the geometry, that is, the size $a$ of the cut in
the $x$-plane (and thus indirectly by the 't Hooft coupling $S$). In
particular one finds the relation $x_1+x_2+x_4=2x_3$. Also, from the
large $x$ behaviour $t\sim x^2$ one derives $A=\hf$.

It will be convenient to shift $t \to t -2c$ with $c=x_3/3$, so that
we can write the relation between $x$ and $t$ as
\eqn\xt{
x= \int^{t-2c} {(t-c) dt\over y}.
}
Here $y(t)$ is given by the canonical Weierstrass form of the elliptic
curve as branched over three point in the $t$-plane
$$
y^2=4t^3-g_2t-g_3.
$$

We recall that the conventional double periodic coordinate $z$ on the
elliptic curve, that satisfies $z \sim z+\w_1$ and $z\sim z+\w_2$ with
modulus $\w_2/\w_1=\tau$, is related to the variables $t$ and $y$
through the Weierstrass $\wp$-function and its derivative
$$
t=\wp(z),\qquad y=\wp'(z),\qquad {dt\over y}=dz.
$$

\ifig\periods{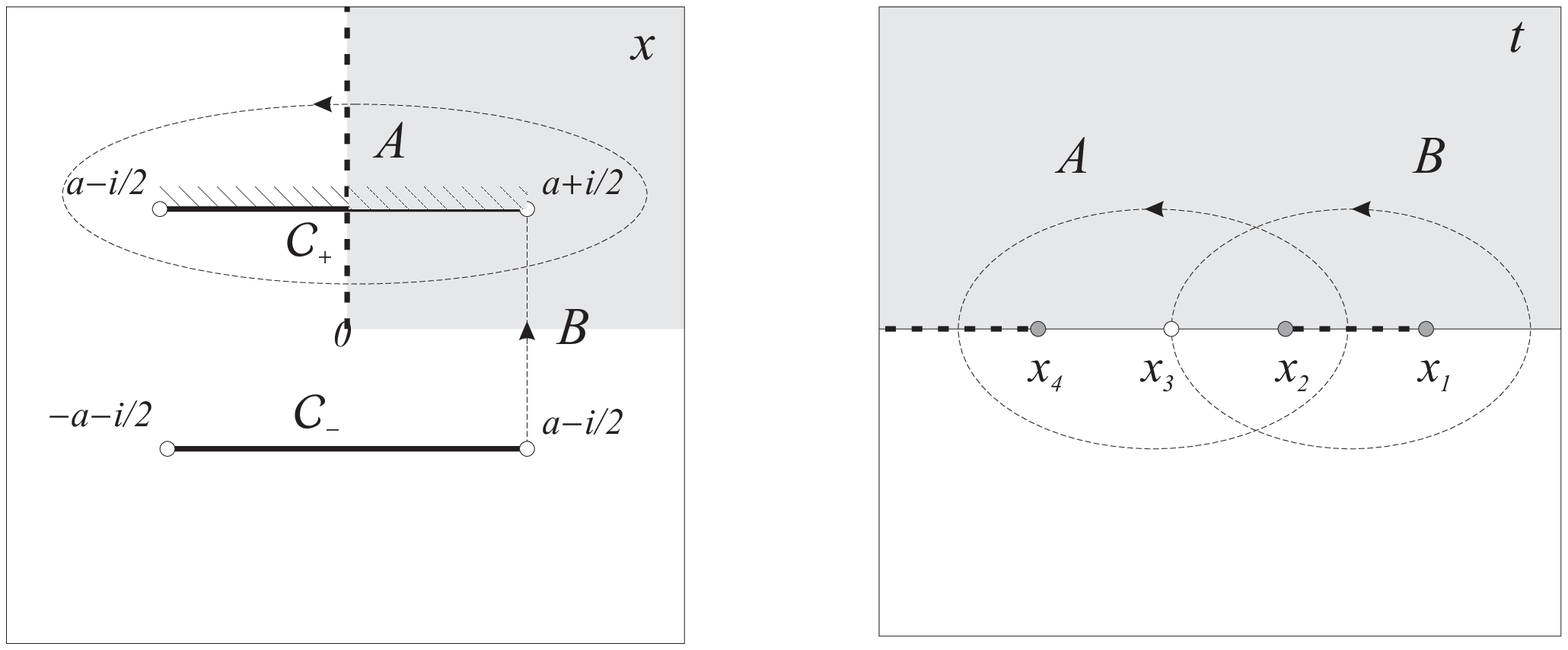}{120}{
The $x$-plane is a double cover of the $t$-plane,
branched at $t=x_1,x_2,x_4$. Note the identifications of the two cuts
in the $x$-plane. The images of the canonical $A$ and $B$ cycles are
indicated.  }

The canonical homology cycles $A$ and $B$ on the torus can now be
described explicitly in the coordinates $x$, $t$, and $z$ (see
\periods). The $A$-cycle encircles clockwise the cut $\cC_+$ in the
$x$-plane (or, after a deformation, counterclockwise the cut
$\cC_-$). In the $t$-plane it encircles the branch points $x_2$ and
$x_4$. Finally in terms of $z$ it runs from $z=0$ to $z=\w_1$.

The dual $B$-cycle is represented in the $x$-plane as the path from
$a-{i\over 2}$ to $a+{i\over 2}$, {\it i.e.}\ as a path that connects
the cut $\cC_-$ to $\cC_+$. In the $t$-plane it encircles the branch
points $x_1$ and $x_2$. It is also the path from $z=0$ to
$z=\w_2=\tau\w_1$.

\subsec{Periods and the superpotential}

Let us now consider what quantities we have to compute to present the
large $N$ solution of the matrix model. First we claim that period of
the one-form $G(x)dx$ along the $A$-cycle measures the total number of
eigenvalues, or more precisely
$$
{1\over 2\pi} \oint_A G(x) dx = g_s N = S.
$$
This follows directly from the fact that the jump in $G(x)$ along the
cut $\cC_+$ (going upwards) is $2\pi S \rho(x)$.

Secondly, the derivative of the planar free energy $\cF_0(S)$ is
obtained by changing the filling number by taking a fraction of
eigenvalues from infinity to the cut. This change is computed as
$$
{\d \cF_0 \over \d S}= \int_\infty^a f(x) dx,
$$
with $f(x)$ the force on the eigenvalue. By using relation
\force\ this can be written as
$$
{\d \cF_0 \over \d S}=  i \int_B G(x)dx.
$$
Originally, the contour along which $G(x)dx$ has to be integrated
starts at the lower cut $\cC_-$, goes to infinity, and then returns
back to the upper cut $\cC_+$. However, since $G(x)$ has no poles at
finite $x$, this contour can be smoothly deformed to the $B$-cycle
that runs directly from $\cC_-$ to $\cC_+$.

In order to perform these period integrals explicitly, we recall the
following standard results. For the $A$-cycles we have
\eqn\Aperiods{
\eqalign{
& \oint_A {dt\over y} = \oint_A dz =\w_1, \cr & \oint_A {tdt\over y} =
\oint_A \wp(z)dz = \eta_1={\pi^2\over 3} E_2(\tau)\w_1\inv, \cr &
\oint_A {t^2 dt\over y} = \oint_A \wp^2(z)dz = {g_2\over 12} \w_1 =
{\pi^4\over 9} E_4(\tau) \w_1^{-3},\cr} 
}
and for the $B$-cycles we find
\eqn\Bperiods{
\eqalign{
& \oint_B {dt\over y} = \oint_B dz =\w_2=\tau \w_1, \cr
& \oint_B {tdt\over y} = \oint_B \wp(z)dz = \eta_2= 
\biggl[{\pi^2\over 3} E_2(\tau) \tau
- 2\pi i\biggr] \w_1\inv,\cr
& \oint_B {t^2 dt\over y} = \oint_B \wp^2(z)dz = {g_2\over 12} \w_2 =
{\pi^4\over 9} E_4(\tau) \tau \w_1^{-3}.\cr}
}
Here $E_2(\tau),E_4(\tau)$ are the standard Eisenstein series, for
example given by the $q$-series expansions ($q=e^{2\pi i \tau}$)
$$
\eqalign{
E_2(\tau) & = 1 -24 \sum_n{nq^n\over 1-q^n},\cr
E_4(\tau) & = 1 +240 \sum_n{n^3 q^n\over 1-q^n}.\cr
}
$$

Returning to the large $N$ solution of our matrix model and in
particular to the periods of the one-form $G(x)dx$, we recall the
relations
$$
G(x)=t+2c,\qquad dx = {(t-c)\over y} dt.
$$
We first compute the periods of the one-form $dx$. The period $\oint_A
dx=0$ gives the relation
$$
c={\pi^2\over 3}E_2 \w_1^{-2}.
$$
The second period $\oint_B dx = i$ fixes the gauge choice
$\w_1=-1/2\pi,$ so that $c=E_2/12$.  Now we are finally in a position
to compute the periods of meromorphic one-form $G(x)dx$. We find
$$
2\pi i S = \Pi_A,\qquad {\d\cF_0 \over \d S} = \Pi_B,
$$
with the periods
$$
\eqalign{
\Pi_A  & = i \oint_A G(x)dx \cr
& = i \oint_A{(t^2+ct-2c^2)\over y} dt \cr
& = {i \pi \over 72}\(E_2^2-E_4\), \cr}
$$
and
$$
\eqalign{
\Pi_B & = i \oint_B G(x)dx  \cr
& =  {i \pi \over 72}\tau \(E_2^2-E_4\) -{1\over 12} E_2.\cr
}
$$
Note the important relation between the two periods
$$
\Pi_B = \tau\,\Pi_A -{1\over 12} E_2(\tau).
$$
The effective superpotential is obtained by inserting these periods in
the expression
$$
W_{\rm eff}(S) = N{\d\cF_0\over \d S}-2\pi i \tau_0\, S = N \Pi_B -
\tau_0\, \Pi_A,
$$
with $\tau_0$ the bare coupling. Now we have to extremize the
superpotential
\eqn\vary{
\delta W_{\rm eff} = N \delta \Pi_B - \tau_0 \, \delta\Pi_A = 0.
}
But the variations of the periods satisfy
\eqn\swgeom{
\delta \Pi_B= \tau\, \delta \Pi_A,
}
with $\tau$ the modulus of the elliptic curve, {\it i.e.}\ we have a
Seiberg-Witten-like geometry. This can be explicitly checked by using the
relation
$$
\delta E_2(\tau) ={i\pi\over 6}\(E_2^2-E_4\) \delta \tau.
$$
Plugging \swgeom\ back in \vary\ we find that
\eqn\tauzero{
\tau = (\tau_0+k)/N,\qquad k=0,1,\ldots,N-1.
}
Here the $\Z_N$ quantum number $k$ distinguishes the different
confining vacua.  So, through the extremization procedure of the
superpotential the modulus $\tau$ of the elliptic curve that appeared
as the master field for the matrix model is identified with $1/N$
times the bare coupling of the gauge theory!  The natural geometric
action of the modular group $SL(2,\Z)$ on $\tau$ in this way gives
precisely the right action predicted by $S$-duality on the coupling
$\tau_0$.

At these critical points the value of the superpotential is given by
$$
W_{\rm eff}= N(\Pi_B-\tau \Pi_A\)
$$ 
with $\tau$ given by \tauzero.  Inserting the values of the periods
$\Pi_A$ and $\Pi_B$, and introducing again the scale $m$ in the
problem by the factor $1/g_s=m^3$ (and matching the overall
normalization to the $\cN=1$ result giving an extra factor $\hf$), we
get the final result
$$
W_{\rm eff}= -{Nm^3\over 24} E_2\({\tau_0+k \over N}\).
$$
in exact agreement with the results of \refs{\dorey,\dk,\ds} (up to an
overall additive constant, but recall that only the differences
$\Delta W_{\rm eff}$ between the vacua are physical). In particular
differentiating with respect to $m$ one gets the one-point function
$$
\<\Tr\F^2\> = -{Nm^2\over 24} E_2\({\tau_0+k \over N}\).
$$
Similarly one can compute the value of the condensate
$$
\< S \> = {m^3\over 4\pi i} \Pi_A = {m^3\over 288}\(E_2^2-E_4\)
=-{1\over 2\pi i}{\d\over \d\tau_0} W_{\rm eff}.
$$

There are various generalizations of this result that we are currently
investigating. For example, we can also consider the Leigh-Strassler
deformation \ls\ of $N=1^*$ given by
$$W(\F)= \Tr\(\F_1[\F_2,\F_3]_\beta + \sum_i m \F_i^2\)$$
where
$$[\F_2,\F_3]_\beta=e^{i \pi \beta}\F_2\F_3-\F_3\F_2e^{-i\pi \beta}.
$$
(The relation of this to non-commutativity in spacetime has been noted
in \bele.)  

The corresponding matrix model with action $W(\F)$ given above
corresponds to a matrix realization of the six-vertex model on random
surfaces \gins\ and has been recently solved in the planar limit in
\kostovsix. Again an elliptic curve features in the solution, as
expected from Montonen-Olive duality.  We can in principal repeat the
analysis we did above for the case at hand, which we shall not
undertake in this paper.

\subsec{$\cN=4$ Yang-Mills on $K3$}

The partition function of $\cN=4$ topologically twisted Yang-Mills
theory was studied on various four manifolds in \vw\ in order to check
the Montonen-Olive conjecture, which predicts that the instanton sum
should give rise to a modular form.  Here we have considered the {\it
untwisted} $\cN=4$ Yang-Mills theory.  However, as noted in \vw\ on
$K3$ the topological twisting is trivial and the topological theory is
equivalent to the untwisted theory.  For simplicity let us only
consider this case, though it is likely that our result applies to all
the Kahler manifolds with $b_2^+>1$ studied in \vw .

As argued in \vw\ all one needs to compute is the correction to the
$\cN=4$ Yang-Mills action proportional to $f(\tau_0) \Tr\,R_+^2$ (more
precisely, including the normalizations, the term ${1\over
2}f(\tau_0)(\chi -{3\over 2}\sigma)$ needs to be computed).  Moreover
for the $SU(2)$ case and for the vacuum with $\F_i=0$ it was found
that
$$f(\tau_0)=-\log\eta(\tau_0/2)$$
As we have argued the same coupling can be determined by summing the
genus one diagrams of the corresponding matrix model.  We have found
from \onch\ that 
$$
f(\tau_0)={\cal F}_1(S)=- \log \eta(\tau),
$$
as the determinant of the chiral boson on the torus is given by the
$\eta$-function.  Using the identification $\tau=\tau_0/N$ and
substituting $N=2$ we reproduce the result of \vw. So in this way one
can try to produce very non-trivial instanton sums by perturbative
means. It would be very interesting to reproduce the results
considered in \vw\ using this approach.

\newsec{Generalizations and conclusions}

Let us summarize the main conclusions of this paper.  We have seen
that chiral $\int \!d^2\theta$ contributions to the effective action
of a $\cN=1$ supersymmetric gauge theory are computed exactly by
summing diagrams of a given genus. In particular the superpotential is
given by the sum of planar diagrams. Similarly the induced
$R^2$-coupling to gravity are given by by genus one
diagrams. Furthermore these diagrams are computed only using the
constant modes and are thus captured by a generalized matrix model.
The exact large $N$ solution of the matrix model, in case such a
solution can be found, will give an effective Calabi-Yau three-fold
geometry. This Calabi-Yau geometry plays the role of the large $N$
master field.  In this way perturbative expansions can be directly
converted to non-perturbative, fractional instanton expansions.

Many dualities, such as Seiberg-like dualities and Montonen-Olive
duality can thus be seen from the planar limit of perturbation theory.
Also Seiberg-Witten geometry can be derived from a purely perturbative
perspective.  It is interesting to note that the cases where there is
an expected $S$-duality from field theory seem to correlate with the
exact solvability of the matrix model, where we can sum up all the
planar diagrams.  In fact the exact sum of the planar diagrams is
needed to see the structure of the $S$-duality group. For example in
the $\cN=1^*$ case modular transformations act on the $A$ and $B$
periods and therefore act essentially as a Legendre transformation on
the superpotential, mixing up the diagrams.  However, even if the
matrix model is not exactly solvable, matrix model perturbation theory
still yields a systematic method to compute instanton corrections to
F-terms in ${\cN}=1$ gauge theories.

Our results lead to many questions for further research. It would be
interesting to study the dimensional reduction to three
dimensions. Could it be that in this case we have to deal with matrix
quantum mechanics; does compactification introduce extra dimensions in
the matrix model, just as in matrix theory?

We have seen that perturbing theories to a massive vacuum, and then
using localization techniques in perturbation theory, is a powerful
way to probe the non-perturbative properties of the field theory. It
would be interesting to apply this philosophy to more general
problems, for example directly in superstring theory.  In particular,
could it be that for a certain class of problems string perturbation
theory has enough information to yield non-perturbative results?

Finally this work might support a long-standing hope, that also in
non-supersymmetric field theories, where we lack the power of
holomorphy, resummations of certain classes of relevant perturbative
diagrams might capture the essential dynamics of the gauge theory.

\bigskip
\centerline{\bf Acknowledgements}

We would like to thank M.~Aganagic, N.~Arkani-Hamed, T.~Banks,
D.~Berenstein, N.~Berkovits, J.~de Boer, F.~Cachazo, N.~Dorey,
M.~Douglas, D.~Gross, M.~Grisaru, S.~Gukov, D.~Kutasov,
K.~Intriligator, R.~Leigh, J.~Maldacena, G.~Moore, N.~Nekrasov,
H.~Ooguri, F.~Quevedo, M.~Rocek, N.~Seiberg, A.~Strominger,
E.~Verlinde, E.~Witten and D.~Zanon for useful discussions and the
organizers of Strings 2002 and the Newton Institute in Cambridge,
where part of this research was done, for providing a most stimulating
research environment. R.D. also wishes to thank the participants of
the 4th Amsterdam Summer Workshop for interesting comments. The
research of R.D.~is partly supported by FOM and the CMPA grant of the
University of Amsterdam, C.V.~is partly supported by NSF grants
PHY-9802709 and DMS-0074329.

\listrefs

\bye